\documentclass[aps, prd, twocolumn, a4paper, 10pt, amsmath, preprintnumbers, nofootinbib, tightenlines, longbibliography, notitlepage,floatfix]{revtex4-2}\pdfoutput=1
\usepackage[english]{babel}
\usepackage{bbm}
\usepackage{graphicx}
\usepackage{hyperref}
\usepackage{wasysym}

\graphicspath{{pics/}}

\newcommand{\ZZ}{\mathbbm{Z}}

\newcommand{\abs}[1]{\lvert#1\rvert}

\newcommand{\cbb}[1]{\big\{#1\big\} }
\newcommand{\cbB}[1]{\Big\{#1\Big\} }

\newcommand{\bh}[1]{\big(#1\big) }
\newcommand{\Bh}[1]{\Big(#1\Big) }

\newcommand{\U}{\Upsilon}
\newcommand{\w}{q}
\newcommand{\e}{\epsilon}

\newcommand{\id}[1]{\operatorname{d}\!#1}
\renewcommand{\d}[2]{\frac{\operatorname{d}\!#1}{\operatorname{d}\!#2}}
\newcommand{\dd}[2]{\frac{\operatorname{d}^2\!#1}{\operatorname{d}\!#2^2}}

\newcommand{\pd}[2]{\frac{\operatorname{\partial}\!#1}{\operatorname{\partial}\!#2}}

\newcommand{\Gc}[2]{G_{#1,#2}}
\newcommand{\Gca}[1]{\underbar{G}_{#1}}
\newcommand{\gca}[1]{\underbar{g}_{#1}}

\newcommand{\gc}[2]{g_{#1,#2}}
\newcommand{\ii}{i}
\newcommand{\ee}{e}

\newcommand{\pt}{\tau}

\newcommand{\mt}{\lambda}

\newcommand{\wa}{\tilde{\w} }
\newcommand{\wap}{\tilde{\w}_\perp }
\newcommand{\Ua}{\tilde{\U} }
\newcommand{\mts}{\tilde\mt}

\newcommand{\ei}{\hat{\e}}
\newcommand{\mti}{\hat\mt}
\newcommand{\wi}{\hat{\w} }
\newcommand{\wip}{\hat{\w}_\perp }
\newcommand{\Ui}{\hat{\U} }

\newcommand{\Uc}{\tilde{\U}^c }
\newcommand{\vUc}{\vec{\tilde{\U}}^c }

\newcommand{\COM}{K}

\newcommand{\nE}{E}
\newcommand{\nL}{L_z}
\newcommand{\nQ}{Q}
\newcommand{\nEd}{\dot{\nE}}
\newcommand{\nLd}{\dot{L}_z}
\newcommand{\nQd}{\dot{\nQ}}

\DeclareMathOperator{\bigO}{O}
\DeclareMathOperator{\sign}{sign}

\newcommand{\ps}{\tilde{p} }

\newcommand{\DV}{\Delta{V}}

\begin{document}

\title{Conditions for sustained orbital resonances in extreme mass ratio inspirals}

\author{Maarten \surname{van de Meent}}
\affiliation{Mathematical Sciences, University of Southampton, Southampton, SO17 1BJ, United Kingdom}
\email{M.vandeMeent@soton.ac.uk}

\date{\today}
\begin{abstract}
We investigate the possibility of sustained orbital resonances in extreme mass ratio inspirals. Using a near-identity averaging transformation, we reduce the equations of motion for a particle moving in Kerr spacetime with self-force corrections in the neighbourhood of a resonant geodesic to a one dimensional equation for a particle moving in an effective potential. From this effective equation we obtain the necessary and sufficient conditions that the self-force needs to satisfy to allow inspiralling orbits to be captured in sustained resonance. Along the way we also obtain the full non-linear expression for the jump in the adiabatic constants of motion incurred as an inspiral transiently evolves through a strong resonance to first-order in the mass ratio. Finally, we find that if the resonance is strong enough to allow capture in sustained resonance, only a small fraction (order of the square root of mass-ratio) of all inspirals will indeed be captured. This makes observation of sustained resonances in extreme mass ratio inspirals---if they exist---very unlikely for space based observatories like eLisa.
\end{abstract} 

\maketitle
\setlength{\parindent}{0pt} 
\setlength{\parskip}{6pt}

\section{Introduction}
An extreme mass ratio inspiral (EMRI) is a binary system consisting of a central supermassive black hole ($\sim{10^6 M_{\astrosun}}$) with a compact object of several solar masses in orbit. As the system emits gravitational radiation, the orbit gradually decays, causing the compact object to slowly spiral towards the central black hole. EMRIs are of great astrophysical interest as a source of gravitational waves for future space-based observatories \cite{Gair:2004iv}. The gradual evolution of EMRIs encodes a very detailed map of the spacetime surrounding the central supermassive black hole. This allows for precise tests of the predictions of general relativity in the strong field regime \cite{Gair:2012nm}. Furthermore, it allows for precise determination of the physical parameters (mass and spin) of the central supermassive black hole \cite{Barack:2003fp}.

The dynamics of EMRIs can be studied by using perturbation theory in the small mass-ratio $\e = m/M \sim 10^{-6}$. At lowest order, the motion of the compact object is that of a test particle following a geodesic in the Kerr spacetime generated by the central massive (and usually rotating) black hole. Any deviations from geodesic motion due to $\e$ taking a finite value, can be included by adding a force term---known as the gravitational self-force---to the geodesic equation. In the last decade a lot of progress has been made in formulating and calculating this self-force (see \cite{Barack:2009ux,Poisson:2011nh} for a comprehensive review).

A common phenomenon in celestial mechanics is the occurrence of resonance, well-known examples include the phase-locking of the rotational and orbital motions of the Moon, and the 3:2 spin-orbit resonance of Mercury. In general, resonances occur when a system has two interacting degrees of freedom oscillating at different frequencies $\omega_1$ and $\omega_2$. If the ratio of these frequencies is rational $\omega_1/\omega_2=n_1/n_2$, then the linear combination $n_2\omega_1-n_1\omega_2$ vanishes. Physically this means that the timescale of the associated interaction terms of the Fourier expansion diverges, leading to behaviour that is qualitatively different from the oscillatory behaviour expected for non-resonant systems.

Geodesic motion in a Kerr background can be characterized by three periods: the time between two successive periastron passes $T_r$, the time between two passes of maximum inclination $T_\theta$, and the time of one sidereal period $T_\phi$. At the geodesic level these oscillations are completely independent. The self-force corrections provide interaction between the oscillations creating the possibility of resonant effects when the periods become commensurate.  Flanagan and Hinderer first worked out the effect explicitly in \cite{Flanagan:2010cd}, although this possibility had been noted in the past.\cite{Mino:2005an,Tanaka:2005ue} Based on a post-Newtonian approximation for the self-force (not necessarily valid in the relativistic strong field regime) they found that as an inspiral system evolves through a situation where $T_r/T_\theta$ is rational, its adiabatic constants of motion (the energy $\nE$,axial angular momentum $\nL$, and Carter constant $\nQ$) acquire a jump of order $\e^{1/2}$.

Over the last few years orbital resonance in EMRIs have gained quite some attention. Grossman, Levin and Perez-Giz \cite{Grossman:2011ps} have studied resonant geodesics in some detail without studying the resonant self-force effects. Gair,  Yunes, and Bender \cite{Gair:2011mr} studied two different toy models for resonance and their effect on the constants of motion. By numerically calculating the ``fluxes'' of the constants of motion to infinity and down the central black hole horizon,  Flanagan, Hughes, and Ruangsri \cite{Flanagan:2012kg} obtained the first true measure of the strength of the resonant effects for realistic relativistic orbits in Kerr spacetime. This was complemented by Ruangsri and Hughes \cite{Ruangsri:2013hra} with a post-Newtonian estimate of the length of each resonant episode, and the time remaining after each episode before the orbiting object plunges into the central black hole. Isoyama et al. \cite{Isoyama:2013yor} studied the adiabatic evolution of the Carter constant when an EMRI system crosses a resonance. Brink, Geyer, and Hinderer \cite{Brink:2013nna} have explored the ``phase space'' of resonant orbits in Kerr spacetime, and noted the relevance of its structure for resonance caused by other astrophysical perturbations that couple the independent oscillations of the geodesic orbits. 

Previous works have focused primarily on so-called \emph{transient resonances}, situations where the EMRI evolves through the resonant situation, with the resonant condition being satisfied only at  one point in time and the duration of the resonance is $\bigO(\e^{-1/2})$. Generally, this is what happens if the terms of the self-force coupling the independent oscillations is weak compared to the terms driving the overall evolution. However, in the general theory of resonant dynamical systems there exists the possibility that the system lingers near the resonance for a prolonged time of order of the inspiral timescale~$\sim \e^{-1}$. Such a situation is known as a \emph{sustained resonance} and its occurrence requires the resonant dynamics to be ``strong'' in some suitable sense. If  a sustained resonance  occurred in an EMRI, it would cause significant qualitative deviation from the normal evolution of the system, which should leave a distinctive imprint on the gravitational waveform.
 
In this paper, we explore exactly what conditions the gravitational self-force needs to satisfy to allow the occurrence of sustained resonances in EMRIs.

\subsection{Background}
Geodesic motion in Kerr spacetime with mass $M=1$ and spin parameter $a$ has four constants of motion, the invariant mass $m$, the energy per unit mass $\nE$, the axial angular momentum per unit mass $\nL$, and Carter's constant  \cite{Carter:1968rr}
\begin{equation}
\begin{split}
\nQ &= \nQ^{\mu\nu}u_\mu u_\nu\\
&=a^2\cos^2\theta (1 -u_t^2)+u_\theta^2+\cot^2\theta u_\phi^2,
\end{split} 
\end{equation} 
where  $\nQ^{\mu\nu}$ is a Killing tensor defined by the second line, $u_\mu$ is the 4-velocity along the geodesic, and $r$, $\theta$, $\phi$, and $t$ are standard Boyer-Lindquist coordinates. These can be used to reduce the geodesic equations from four second-order differential equations to four first-order equations,
\begin{subequations}\label{eq:gd}
\begin{align}
\Bh{\d{r}{\pt}}^2 &=\frac{ R(r)}{\Sigma(r,\theta)^2}, \\
\Bh{\d{\theta}{\pt}}^2 &= \frac{\Theta(\cos\theta)}{\Sigma(r,\theta)^2},  \\
\d{\phi}{\pt} &=\frac{\Phi_r(r)+\Phi_\theta(\theta)}{\Sigma(r,\theta)},  \\
\d{r}{\pt} &= \frac{T_r(r)+T_\theta(\theta)}{\Sigma(r,\theta)},  
\end{align} 
\end{subequations} 
where $\pt$ is proper time, $R$, $\Theta$, $\Phi_j$, and $T_j$ are functions of $r$ or $\theta$, which can be obtained explicitly in terms of $\nE$, $\nL$, and $\nQ$, and $\Sigma=  r^2+a^2\cos^2\theta$. As first noted by Carter \cite{Carter:1968rr}, and more recently emphasized by Mino,\cite{Mino:2003yg} these equations can be simplified by choosing an alternative time parametrization~$\mt$ defined by,
\begin{equation}
 \d{\pt}{\mt}=\Sigma =  r^2+a^2\cos^2\theta.
\end{equation} 
This time parameter is commonly referred to as ``Mino time''. The effect of this choice is that all $\Sigma$'s disappear from \eqref{eq:gd}, and the equations for $r$ and $\theta$ decouple.

Since the geodesic equations in Kerr have as many constants of motion as equations (and their mutual Poisson brackets vanish), the system of equations is integrable \cite{Hinderer:2008dm}. One interesting property (out of many) of integrable systems is that they can be rewritten in action-angle variables $(\w^\mu,J^\mu)$ as
\begin{subequations}
\begin{align}
 \d{\w_\mu}{\mt} &= \U_\mu(J),\\
  \d{J_\mu}{\mt} &= 0. 
\end{align} 
\end{subequations} 
Schmidt \cite{Schmidt:2002qk} used this property to obtain the frequencies of Kerr orbits with respect to Boyer-Lindquist coordinate time $\Omega_\mu$, while analytic expressions for the frequencies with respect to Mino time $\U_\mu$ were obtained by Fujita and Hikida \cite{Fujita:2009bp}.

\begin{figure}[tbp]
\includegraphics[width=\columnwidth]{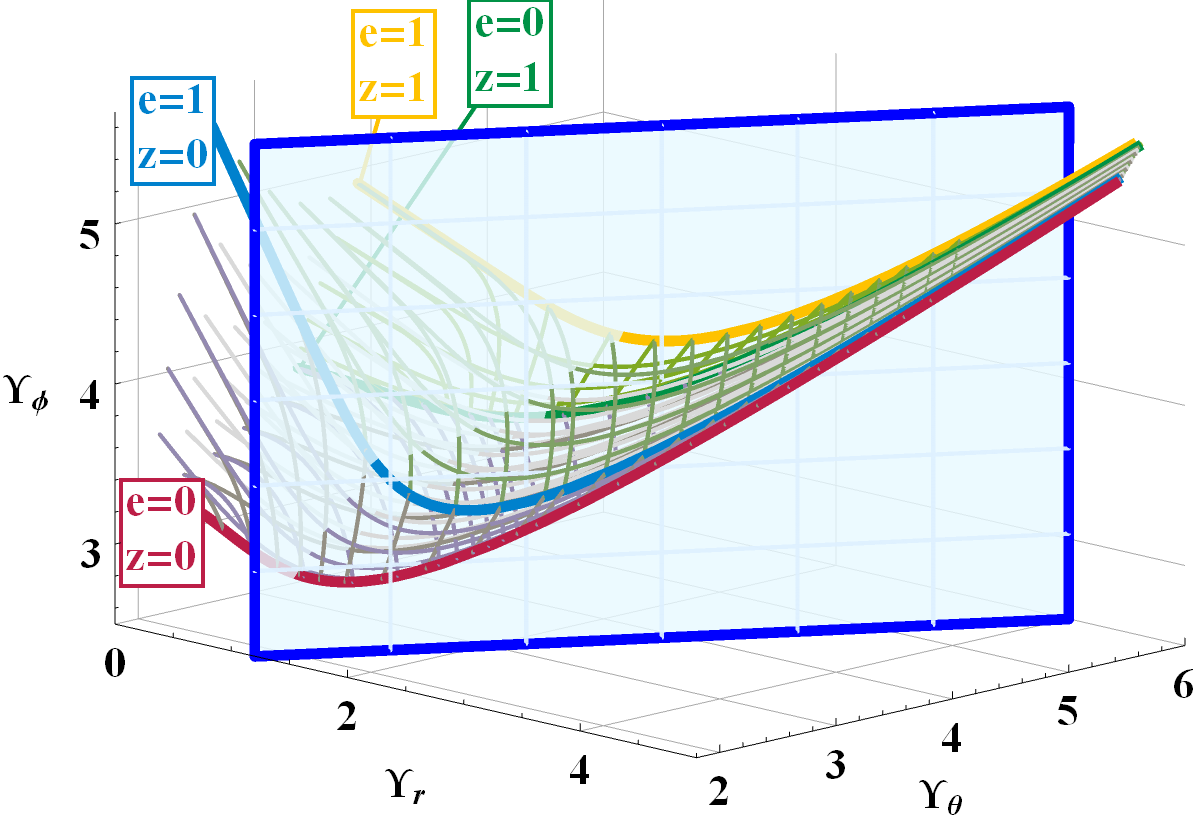}
\caption{A plot of parameter space of bound geodesic in a Kerr space time with $a=0.9$. The plot curves are lines of constant semi-latus rectum $p$, eccentricity $e$, and/or $z_{max}=\max_{\lambda} \cos\theta(\lambda)$. The labelled lines represent equatorial circular, equatorial parabolic, polar circular, and polar parabolic orbits. The transparent surface represents all orbits in a 3:2 resonance. (See appendix \ref{app:Minofreq} for more details.)}\label{fig:phasediag}
\end{figure}

Figure \ref{fig:phasediag} shows the parameter space of orbits in Kerr in terms of the frequencies with respect to Mino time. There are some features that are worth commenting on here. First, due to the equivalence principle the orbital dynamics cannot depend on the mass of the orbiting object $m$, hence we only need three parameters (e.g. $\U_r$, $\U_\theta$, and $\U_\phi$)\footnote{See appendix \ref{app:Minofreq} for discussion on the validness of $\U_r$, $\U_\theta$, and $\U_\phi$ as parameters on the space of bound orbits.} to describe each orbit. Second, the Mino frequencies all diverge as the radius of the orbit increases---in sharp contrast to the frequencies with respect to coordinate time or proper time which vanish at infinity. This is a result of the fact that Mino time  is rescaled by a factor $\Sigma= r^2 + a^2\cos^2\theta$ with respect to proper time. Third, as the radius increases the three Mino time frequencies converge to a common (diverging) value.  Fourth,  at the separatrix  dividing bound orbits and orbits plunging into the central black hole $\U_r$ is zero.

A resonant orbit is a geodesic for which the ratio $\U_r/\U_\theta$ is a rational number, or equivalently, for which there exist coprime integers $n_r$ and $n_\theta$ such that
\begin{equation}\label{eq:rescond}
n_r \U_r + n_\theta\U_\theta =0.
\end{equation} 
The sum $\abs{n_r}+\abs{n_\theta}$ is called the \emph{order} of the resonance. As a rule of thumb for resonant dynamical systems low order resonances have greater effect than high order resonance \cite{Flanagan:2010cd}. Hence, we will generally focus on the low order resonances. The resonant condition \eqref{eq:rescond} defines a two-dimensional subspace of the parameter space, which we refer to as a \emph{resonant surface}. In figure \ref{fig:phasediag} we have plotted the resonant surface for the 3:2 resonance.

Note that at infinity the ratio $\U_r/\U_\theta$ is unity, while at the separatrix this ratio is zero. Hence, as an inspiral evolves from infinity to a plunge into the central black hole it passes through all resonant surfaces with $\abs{n_r} > \abs{n_\theta}$. Resonances are therefore a generic feature of any EMRI.

\subsection{Overview of this paper}
In this paper we will derive the exact conditions that the self-force needs to satisfy to allow sustained resonances to occur in EMRIs. In section \ref{sec:eom} we first review how the self-force can be re-expressed as a correction to the equations of motion for $\U_i$ and $\w_\mu$, then we apply a near identity transformation to simplify these equations of motion by absorbing all non-resonant interaction terms in $\U_i$ and~$\w_\mu$.

In section \ref{sec:res} we then discuss the effect of resonant interactions. We start in section \ref{sec:effpot} with deriving a set of effective equations of motion in the neighbourhood of a resonant surface. At leading-order this equation describes the conservative dynamics of a particle moving in a one-dimensional effective potential. In section \ref{sec:trans} we then use this effective potential to derive the jump in the constants of motion induced by a transient crossing of a resonant surface. Section \ref{sec:sust} then continues to derive a  necessary and sufficient condition for the existence of sustained resonance solutions to the equation of motion, which is related to the existence of local minima in the effective potential. Existence of such solutions however is not sufficient to conclude that they also occur in EMRI. This requires that an inspiral starting far away from the black hole is captured in a local minimum of the potential. Section \ref{sec:capt} derives the necessary and sufficient condition for this to happen as a restriction on the initial conditions. Finally, section \ref{sec:escape} discusses the fate of an EMRI after it is captured in sustained resonance.

The final section (\ref{sec:test}) then discusses  how the above conditions can be tested using numerical calculations of the self-force, and what can be learned from already existing numerical results.

\subsection{Notations and Conventions}
We employ units such that $c$ (speed of light) $G$ (Newton's constant), and $M$ (the mass of the central supermassive black hole) are all unity. Consequently, all quantities appearing in this paper are dimensionless. Furthermore the constants of motion $\nE$, $\nL$, and $\nQ$ are normalized to be independent of the invariant particle mass $m$. Metrics have signature  $(- + + +)$. Without further specification an index $i$ runs over $(r,\theta,\phi)$, and an index $j$ runs over $(r,\theta)$. Repeated indices are generally summed over their full range, unless otherwise indicated. The Mino frequencies $\U_r$ and $\U_\theta$ are considered positive by convention, $\U_\phi$ can be either positive or negative with $a\U_\phi$ positive for prograde orbits and  $a\U_\phi$ negative for retrograde orbits (where $a$ is the spin parameter of the central supermassive black hole).

\section{Equations of motion}\label{sec:eom}
\subsection{First-order equations with self-force}
In the limit $\e=0$ the orbiting object follows a geodesic of the background spacetime. The self-force program summarizes corrections order-by-order in $\e$ as a term on the right-hand side (RHS) of the geodesic equation,
\begin{equation}
\dd{x^\nu}{\pt}+\Gamma^{\nu}_{\sigma\rho}\d{x^\sigma}{\pt}\d{x^\rho}{\pt} =\e a^\mu(x),
\end{equation} 
where $\e a^\mu$ is the self-acceleration. Hinderer and Flanagan show in \cite{Hinderer:2008dm} how this second-order equation can be rewritten as two first-order equations,
\begin{subequations}\label{eq:eomAA}
\begin{align}
\d{P_i}{\mt} &= \e \tilde{G}_i(\vec{P},\w_r,\w_\theta) + \bigO(\e^2),\\
\d{\w_\nu}{\mt} &= \U_\mu(\vec{P}) +  \e\tilde{g}_\nu(\vec{P},\w_r,\w_\theta) + \bigO(\e^2),
\end{align}
\end{subequations}
where the $\w_\mu$ are generalized angle variables, and  $\vec{P}=(P_1,P_2,P_3)=(\nE,\nL,\nQ)$ are the slowly varying constants of motion. In particular they show how the forcing terms $\tilde{G}_i$ and $\tilde{g}_\nu$ follow from the self-acceleration $a^\nu$ (see appendix \ref{app:SF} for an explicit formula).

The frequencies $\U_\nu$ themselves are functions of the $P_i$, and therefore constants of motion of the zeroth order system. We can therefore simplify the analysis of the system by using $\vec\U = (\U_r,\U_\theta,\U_\phi)$ as the slow variables instead of $\vec{P}$. One might worry that the map $\vec{P} \mapsto \vec{\U}(\vec{P})$ is not invertible due to the existence of isofrequency pairs. It is known that this is the case for the frequencies $\Omega_i$ with respect to Boyer-Lindquist coordinate time \cite{Warburton:2013yj}. In appendix \ref{app:Minofreq} we present evidence that no such pairing occurs for  frequencies with respect to Mino time. In the rest of this paper we will assume that this map is indeed invertible, and use $\vec\U$ as slow variables. However, the results do not crucially depend on this assumption.

Note that the RHS of equations \eqref{eq:eomAA} does not depend on $\w_t$ and $\w_\phi$. The equations for these two generalized angles thus decouple, and we can focus on just $\w_r$ and $\w_\theta$ as the fast variables, since  $\w_t$ and $\w_\phi$ can later be recovered by a simple integration. This yields the reduced system,
\begin{subequations}\label{eq:eom}
\begin{align}
\d{\U_i}{\mt} &= \e G_i(\vec{\U},\vec{\w}) + \bigO(\e^2),\\
\d{\w_j}{\mt} &= \U_j +  \e g_j(\vec{\U},\vec{\w}) + \bigO(\e^2),
\end{align}
\end{subequations} 
where $\vec\w=(\w_r,\w_\theta)$, $i$ runs over $(r,\theta,\phi)$ and $j$ runs over $(r,\theta)$ . Explicit formula for obtaining $G_i$ and $g_j$ can found in \cite{Hinderer:2008dm} or appendix \ref{app:SF}.

\subsection{Averaged equations near resonance}
Since the inspiral timescale of the system \eqref{eq:eom},  $T_{insp} = \bigO(\e^{-1})$, is much larger than the orbital timescale $T_{orb}= \bigO(1)$, one intuitively expects the solutions to be largely independent of the rapid oscillations of the forcing functions $G_i$ and $g_j$, and to mainly depend on their averages. This intuition fails near a resonance, where the period of some of the oscillating terms diverges. However, one still expects the other oscillating terms whose periods stay small with respect to the inspiral timescale, to play only a subdominant role. In this section we describe how to make this notion precise by employing a so-called near-identity averaging transformation following \cite{KC:1996}.

Suppose that the system encounters a resonance at $\mt=0$, i.e. for some coprime integers $n_r$ and $n_\theta$  the linear combination $n_r\U_r+n_\theta\U_\theta\equiv\U_\perp$ vanishes at $\mt=0$.\footnote{Note that there is a sign ambiguity in the definition of $\U_\perp$. We will adopt the convention that  $n_r>0$ and $n_\theta<0$. This has the consequence that in an inspiral $\U_\perp$ starts out positive and becomes negative as the infalling object crosses the resonance. }

Since the generalized angles $\w_r$ and $\w_\theta$ have a period of $2\pi$, the functions $G_i$ and $g_j$ can be decomposed as Fourier series,
\begin{subequations}\label{eq:fourierexp}
\begin{align}
\begin{split}
G_i(\vec\U,\vec{\w}) &= \Gca{i}(\vec\U) + \sum_{N\neq 0} \Gc{i}{N}(\vec\U)\ee^{\ii N \w_\perp}\\ 
&\quad\qquad+\sum_{(n,k)\in R} \Gc{i}{nk}(\vec\U) \ee^{\ii n \w_r+\ii k \w_\theta},
\end{split} \\
\begin{split}
g_j(\vec\U,\vec{\w}) &= \gca{j}(\vec\U) + \sum_{N\neq 0} \gc{j}{N}(\vec\U)\ee^{\ii N \w_\perp}\\ 
&\quad\qquad+\sum_{(n,k)\in R} \gc{j}{nk}(\vec\U) \ee^{\ii n \w_r+\ii k \w_\theta},
\end{split} 
\end{align} 
\end{subequations}
where we introduced the resonant phase combination $\w_\perp = n_r \w_r +n_\theta \w_\theta$,\footnote{In the rest of this paper whenever we have a quantity $X_i$ with $i$ running over $(r,\theta,\phi)$ or $(r,\theta)$, we denote the linear combination $n_r X_r +n_\theta X_\theta$ by $X_\perp$.} the sum over $N$ runs from $-\infty$ to $\infty$ , and $R$ is the set $\{\left (n,k)\in\ZZ | (n,k) \neq N(n_r,n_\theta), \forall N\in\ZZ\right\}$ of all non-resonant 2-tuples. The first term in the expansion of \eqref{eq:fourierexp} is the average adiabatic term, the first sum contains all the resonant terms, i.e. all oscillating terms whose period diverges at the resonance, and the final sum contains all the non-resonant oscillating terms.

We employ an appropriate near-identity averaging transformation (see \cite{KC:1996} sec 5.1.3),
\begin{subequations}
\begin{align}
\Ua_i &= \U_i +\e T_i(\vec\U,\vec{\w}) +\bigO(\e^2),\\
\wa_j &= \w_j +\e L_j(\vec\U,\vec{\w}) +\bigO(\e^2),
\end{align} 
\end{subequations} 
where $T$ and $L$ are functions of $\vec\U$ and $\vec{\w}$ defining the transformation. The goal is to choose  $T$ and $L$  in such a way as to eliminate the dependence of  \eqref{eq:eom} on the rapidly oscillating non-resonant phase can be eliminated to order $\e$. The details of this transformation are described in appendix \ref{app:niat}. The result is,
\begin{subequations}\label{eq:eomavg}
\begin{align}
\d{\Ua_i}{\mt} &= \e G_i(\vec\Ua,\wap)+ \bigO(\e^2),\\
\d{\wa_j}{\mt} &= \Ua_j +\e g_j(\vec\Ua,\wap)  + \bigO(\e^2),
\end{align} 
\end{subequations} 
with
\begin{subequations}
\begin{align}
G_i(\vec\Ua,\wap) &= \Gca{i}(\vec\Ua) +\sum_{N\neq 0} \Gc{i}{N}(\vec\Ua)\ee^{\ii N \wa_\perp},\\
g_j(\vec\Ua,\wap) &=\sum_{N\neq 0} \gc{j}{N}(\vec\Ua)\ee^{\ii N \wa_\perp}.
\end{align} 
\end{subequations} 
An important consequence of this result is that the forcing terms now only depend on the resonant phase $\wip$. The equations of motion for the other phases thus decouple from the system, and we can first focus on solving the system for $\w_\perp$ and $\vec\Ua$. The other phases can then be recovered by direct integration.

\section{Resonances}\label{sec:res}
\subsection{Effective potential}\label{sec:effpot}
In order to solve the averaged equations of motion \eqref{eq:eomavg} near  the resonance at $\mt=0$ we introduce a rescaled boundary layer timescale
\begin{equation}
 \mti = \ei\mt,
\end{equation} 
with $\ei\equiv\e^\xi$ for some power $\xi>0$. By expanding $\Ua_i$ and $\wa_\perp$ in $\ei$,
\begin{subequations}\label{eq:transexp}
\begin{align}
\Ua_i(\mt,\e) &= \Ua_i^0+\ei \Ui^1_i(\mti) + \bigO(\ei^2),\\
\wa_\perp(\mt,\e) &=\wi^0_\perp(\mti) +\ei \wi^1_\perp(\mti) + \bigO(\ei^2).
\end{align} 
\end{subequations} 
and plugging these into the averaged equations of motion \eqref{eq:eomavg} we obtain at leading-order
\begin{subequations}\label{eq:eomtrans}
\begin{align}
\d{\Ui^1_i}{\mti} &=\ei^\frac{1-2\xi}{\xi}G_i(\vec\Ua^0,\wip^0),\label{eq:eomtransfreq}\\
\d{\wi^0_\perp}{\mti} &= 
\Ui^1_\perp.\label{eq:eomtransphase}
\end{align} 
\end{subequations}
Obtaining a non-singular and non-trivial limit as $\ei\to 0$ requires that we set $\xi=1/2$.

Differentiating \eqref{eq:eomtransphase} with respect to $\mti$ and plugging in  \eqref{eq:eomtransfreq} gives the second order equation,
\begin{equation}\label{eq:eom2nd}
\dd{\wip^0}{\mti}=G_\perp(\vec\Ua^0,\wip^0).
\end{equation}
By multiplying both sides of this equation with $\d{\wip^0}{\mti}$ and integrating once with respect to $\mti$, we obtain an equivalent first-order equation,
\begin{equation}\label{eq:eom1st}
\frac{1}{2}\Bh{\d{\wi^0_\perp}{\mti}}^2= K-V(\wip^0),
\end{equation}
where 
\begin{equation}
V(\w)= - \Gca{\perp}(\vec\Ua^0)\w +\ii \sum_{N\neq 0} \frac{\Gc{\perp}{N}(\vec\Ua^0)}{N}\ee^{\ii N \w},
\end{equation} 
and $\COM$ is an integration constant.

\begin{figure}[tbhp]
\includegraphics[width=\columnwidth]{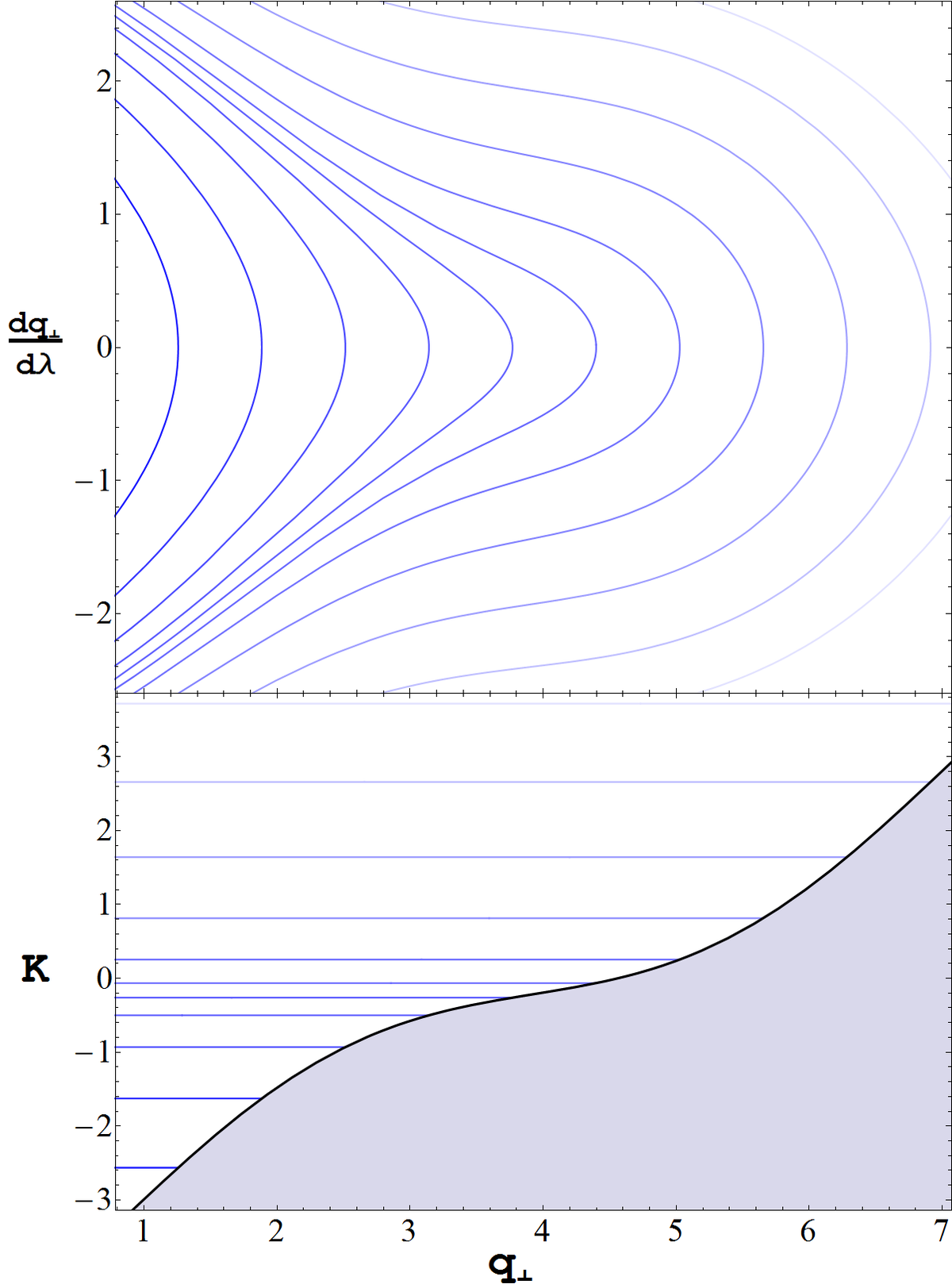}
\caption{The top figure shows the phase portrait of solutions of \eqref{eq:eom1st}, with $V=\wip-\frac{1}{2}\cos\wip+\frac{1}{2}\sin\wip$. The bottom figure show the potential of the same system. The levels plotted in the lower figure correspond to the solutions plotted in the top figure with the same shade.}\label{fig:trans}
\end{figure}

This is the phase resonance equation studied in \cite{Gair:2011mr} as a toy model. It can be interpreted as the equation for a 1D particle moving in an effective potential, $V$ with energy $\COM$. Figure \ref{fig:trans} plots the solutions of this equation in the case that $V=\wip-\frac{1}{2}\cos\wip+\frac{1}{2}\sin\wip$.

\subsection{Transient resonance}\label{sec:trans}
As discussed in \cite{Flanagan:2010cd}, the effect of a resonance compared to the adiabatic approximation ignoring the oscillating terms is an order $\ei$ jump in the slow variables $\vec\Ua$. We now provide an expression for the size of these jumps. 

The initial condition that the system crosses the resonance at $\mt=\mti=0$ corresponds to setting,
\begin{equation}
K = V(\w_0),
\end{equation}
where $\w_0$ is the value of $\wip(\mti)$ at $\mti=0$. We can then invert \eqref{eq:eom1st} to find $\mti$ as a function of $\wip$.
\begin{equation}\label{eq:lamint}
\mti(\wip) =\int_{\w_0}^{\wip}\frac{1}{\pm\sqrt{2(\COM-V(\w))}} \id{\w},
\end{equation}
where  the $\pm$ sign corresponds to the branch before and after resonance.

The effective potential $V$ can be split into an adiabatic contribution $\underbar{V}$ and a (resonant) oscillatory contribution $\DV$,
\begin{subequations}
\begin{align}
V(\wip) &= \underbar{V}(\wip)+\DV(\wip),\\
\underbar{V}(\wip) &=  - \Gca{\perp}(\vec\Ua^0)\wip,\\
\DV(\wip) &= \ii \sum_{N\neq 0} \frac{\Gc{\perp}{N}(\vec\Ua^0)}{N}\ee^{\ii N \wip}.
\end{align} 
\end{subequations} 
In the adiabatic approximation (i.e. ignoring $\DV$) equation \eqref{eq:lamint} can be integrated explicitly, yielding
\begin{equation}
\mti(\wip) = \pm 2 \sqrt{\frac{\w_0-\wip}{-\Gca{\perp}}}.
\end{equation} 

If we denote the full integral \eqref{eq:lamint} by $\mti$ and its adiabatic approximation $\mti_0$, then as the system ``evolves'' from $\wip=-\infty$ to $\w_0$ to $-\infty$ again the full integral ``accumulates'' an additional amount of time over the adiabatic approximation given by,
\begin{equation}\label{eq:mtjump}
\begin{split}
\!\Delta\mti &= \lim_{\w\to-\infty} \mti(\w)- \mti_0(\w)+ \lim_{\w\to\infty} \mti(\w)- \mti_0(\w)\\
&=\!\int_{-\infty}^{\w_0}\!\!\big(\frac{\sqrt{2}}{\sqrt{V(q_0)-V(\w)}}- \frac{\sqrt{2}}{\sqrt{\underbar{V}(q_0)-\underbar{V}(\w)}}\big)\id{\w}.
\end{split}
\end{equation}
In the asymptotic regime $1/\ei\gg\abs{\mti}\gg 1$, far away from the resonance the resonant terms $\DV$ are oscillating rapidly and can be ignored at lowest order. Equation \eqref{eq:eom2nd} then has solutions of the form
\begin{equation}
\wip(\mti) = \frac{\Gca{\perp}}{2} \Big(\mti+\sign(\mti)\frac{\Delta\mti}{2}\Big)^2 + C,
\end{equation} 
where $C$ is some constant that is not relevant here. From equation \eqref{eq:eomavg}, we then find the total jump in the frequencies $\Ua$,
\begin{equation}\label{eq:Ujump}
\Delta\Ua_i = \ei\Gca{i}\Delta\mti +\bigO(\ei^2).
\end{equation} 
Together, equations \eqref{eq:mtjump} and \eqref{eq:Ujump} provide a full expression for this jump (to leading-order in $\ei$).\footnote{This result depends on the assumption that the resonance is transient, consequently it does not apply in the sustained resonance situations discussed in the next sections.} In previous works \cite{Gair:2011mr,Flanagan:2010cd}, the size of this jump was only calculated in the limit that $\abs{\DV}\ll \abs{\Gca{\perp}}$ (although a similar result is supposed to appear in \cite{FH:??}). To check our result with theirs, we expand  \eqref{eq:mtjump} in $\DV$,
\begin{subequations}
\begin{align}
\Delta\mti &\approx -\frac{1}{\sqrt{2}}\int_{-\infty}^{\w_0} \frac{\DV(\w_0)-\DV(\w)}{(\underbar{V}(\w_0)-\underbar{V}(\w))^{3/2}}\id{\w},\\
&= -\ii\sum_{N\neq 0}\frac{\Gc{\perp}{N} }{\sqrt{2}N\Gca{\perp}^{3/2}}
\int_{-\infty}^{q_0} \frac{\ee^{\ii N\w_0}-\ee^{\ii N\w}}{(\w_0-\w)^{3/2}}\id{\w}.
\end{align} 

\end{subequations} 
This integral can be computed explicitly,
\begin{equation} 
\Delta\mti = \sum_{N\neq0} \frac{\sqrt{2\pi}}{\abs{\Gca{\perp}N}^{\frac{1}{2}}}
\frac{\Gc{\perp}{N}}{\Gca{\perp}}
\ee^{\ii N \w_0-\ii\frac{\pi}{4}\sigma},
\end{equation}
where $\sigma$ is the sign of $N$. Plugging this result into equation \eqref{eq:Ujump} yields,
\begin{equation} 
\Delta\Ua_i = \ei\Gca{i}
 \sum_{N\neq0} \frac{\sqrt{2\pi}}{\abs{\Gca{\perp}N}^{\frac{1}{2}}}
\frac{\Gc{\perp}{N}}{\Gca{\perp}}
\ee^{\ii N \w_0-\ii\frac{\pi}{4}\sigma},
\end{equation} 
which agrees with the linear result in \cite{Flanagan:2010cd}.

\subsection{Sustained resonance}\label{sec:sust}
\begin{figure}[tb]
\includegraphics[width=\columnwidth]{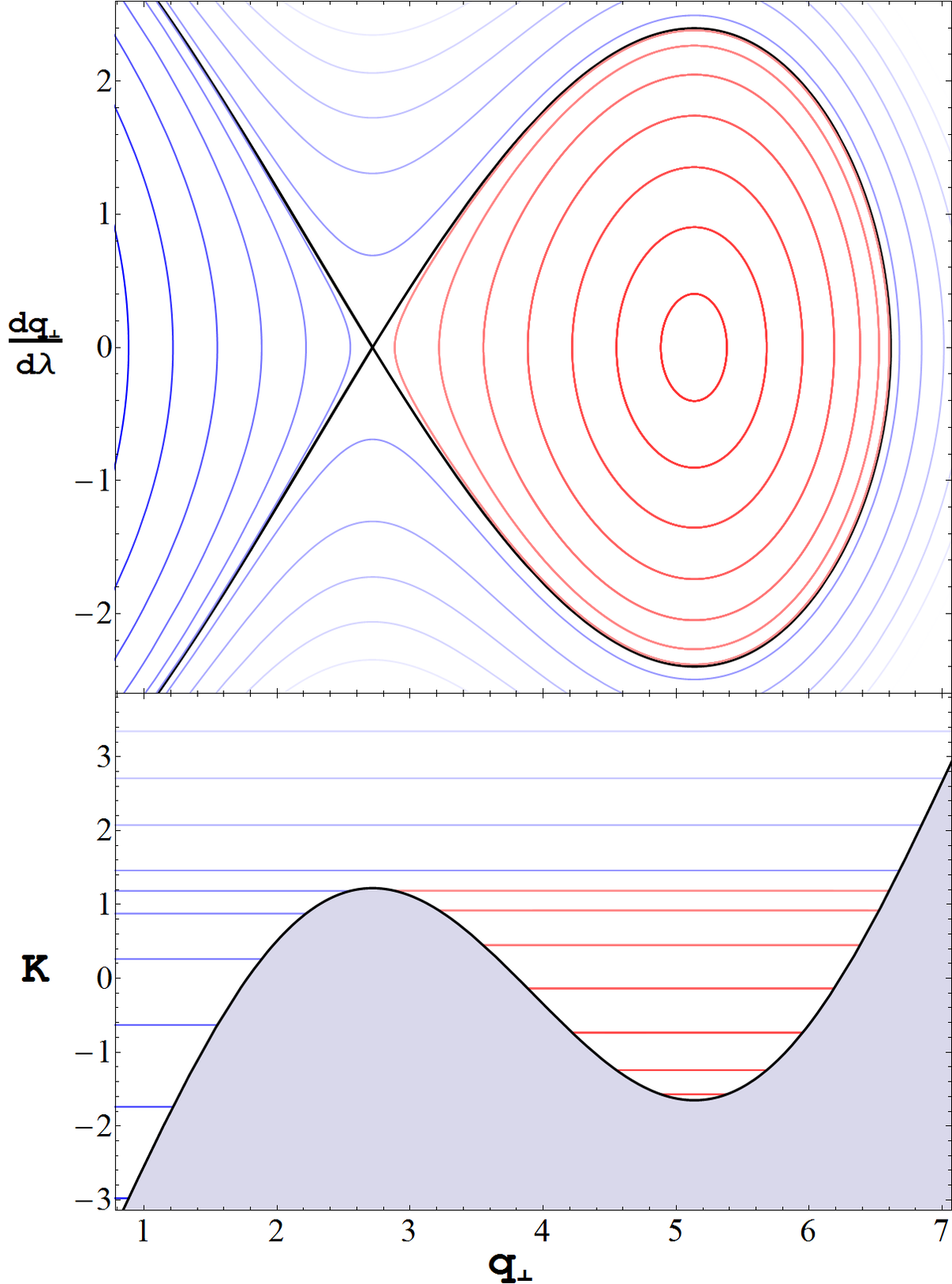}
\caption{Plots of the same system as in figure \ref{fig:trans}, but with $V=\wip-2\cos\wip+2\sin\wip$. The phase portrait now has two fixed points related to the local minimum (stable) and maximum (unstable) of the potential. Near the stable fixed point there exist periodic solutions of the system (in red) that stay near the fixed point.}\label{fig:sust}
\end{figure} 
When the effective potential $V$ has a local minimum, equation \eqref{eq:eom1st} has solutions that oscillate around the resonant surface (see figure \ref{fig:sust}). Such solutions are called \emph{sustained resonances}. As time progresses, the system will oscillate around the resonance surface, and the components of $\vec\Ui^1$ perpendicular to $\Ui^1_\perp$ will continue to grow. Eventually, they become order $\ei^{-1}$ and the expansion \eqref{eq:transexp} becomes disordered.

To follow the long term evolution of such a sustained resonance it is convenient to introduce the rescaled orbital parameters $\Ui_i(\mti,\ei)$  and $\wi(\mti,\ei)$,
\begin{subequations}
\begin{align}
\Ua_i(\mt,\e) &= \Uc_i(\mts)+\ei \Ui_i(\mti,\ei),\\
\wa(\mt,\e) &=\wi(\mti,\ei),
\end{align} 
\end{subequations} 
where  $\Uc_i(\mts)$ is an a priori unknown set of functions of the slow time $\mts = \ei \mti = \e \mt$, satisfying  $\Uc_\perp(\mts)=0$, which describes the slow evolution of the system along the resonant surface.

Plugging these into \eqref{eq:eomavg} and expanding in $\ei$ yields
\begin{subequations}
\begin{align}\label{eq:eomsust}
\begin{split}
\d{\Ui_i}{\mti} &=-\d{\Uc_i}{\mts}+
\Gca{i} + \sum_{N\neq 0} \Gc{i}{N}\ee^{\ii N \wip}+\\
&\qquad\ei \Ui_j \big(\d{\Gca{i}}{\U_j}+\sum_{N\neq 0}   \d{\Gc{i}{N}}{\U_j}\ee^{\ii N \wip}\big)+ \bigO(\ei^2),
\end{split} \\
\d{\wip}{\mti} &= \Ui_\perp + \ei\sum_{N\neq 0} \gc{i}{N}\ee^{\ii N \wip} + \bigO(\ei^2),
\end{align} 
\end{subequations} 
where summation over the repeated index $j$ is implicit and the functions $G$ and $g$ and their derivatives are to be evaluated at $\vUc$.
Bosley and Kevorkian in \cite{Bosley:1995} describe how to solve such equations to find the evolution of a system captured in sustained resonance. For our current purpose it is enough to note such sustained resonance solutions can exist in principle and stay near the resonance for a prolonged period resulting in a qualitatively different behaviour from the adiabatic approximation of an inspiral.

A necessary and sufficient condition for sustained resonance solutions to exist is that the effective potential for the resonant phase $V$ has a local minimum.  I.e. there must be values of the resonant phase $\wip$ for which $\d{V}{\wip}=0$.\footnote{Strictly speaking, this only implies the existence of stationary points. However, the periodic nature of $V'$ in $\wip$ ensures that local mimima and maxima must come in pairs.} For the Fourier components of the self-force this means
\begin{equation}
 \Gca{\perp}(\vec\Ua^0) = -\sum_{N\neq 0}\Gc{\perp}{N}(\vec\Ua^0)\ee^{\ii N \wip}.
\end{equation} 
Since the left hand side is constant and the right hand side is purely oscillatory, this equation only has solutions if the right hand side has an amplitude which is bigger than  $\abs{ \Gca{\perp}(\vec\Ua^0)}$, i.e. 
\begin{equation}\label{eq:deltaG}
\Delta G_\perp(\vec\Ua^0)\equiv\frac{G^{max}_\perp(\vec\Ua^0)-G^{min}_\perp(\vec\Ua^0)}{\abs{G^{max}_\perp(\vec\Ua^0)+G^{min}_\perp(\vec\Ua^0)}}\geq 1,
\end{equation} 
where
\begin{align}
G^{max}_\perp(\vec\Ua) &= \max_{\w_\perp\in[0,2\pi]} G_\perp(\vec\Ua,\w_\perp),\\
G^{min}_\perp(\vec\Ua) &= \min_{\w_\perp\in[0,2\pi]} G_\perp(\vec\Ua,\w_\perp).
\end{align} 
In terms of the Fourier components of the self-force this implies the necessary (but not sufficient) condition,
\begin{equation}
 \abs{\Gca{\perp}(\vec\Ua^0)}< \sum_{N\neq 0}\abs{\Gc{\perp}{N}(\vec\Ua^0)},
\end{equation} 
for the existence of sustained resonances. The negation of this condition is therefore sufficient (but not necessary) to show that there are no sustained resonances. 

\subsection{Capture condition}\label{sec:capt}
However, the existence of sustained resonance solutions to equation \eqref{eq:eomavg}, does not guarantee that they will also occur for inspiralling solutions. In fact, in the lowest order approximation of this equation given by \eqref{eq:eom1st} $\COM$ is a constant of motion; consequently, if a solution starts at $\w_\perp=-\infty$ it cannot get stuck in a local potential well (see figure \ref{fig:sust}), and will in general return to $\w_\perp=-\infty$ after reflecting off the potential. The only exceptions are the isolated solutions that asymptote to the unstable equilibrium at a local maximum of the potential. 

In this section we derive the conditions under which  EMRI may be captured in sustained resonance. In \cite{Haberman:1983} Haberman derived these conditions for a Hamiltonian system with one degree of freedom, where the resonance is caused by a perturbation that is a predetermined function of time. We generalize his method to apply to orbital resonances in Kerr spacetime, where there are three slowly evolving frequencies and the perturbation causing the resonance (the self-force) is a function of the frequencies.

\begin{figure}[tb]
\includegraphics[width=\columnwidth]{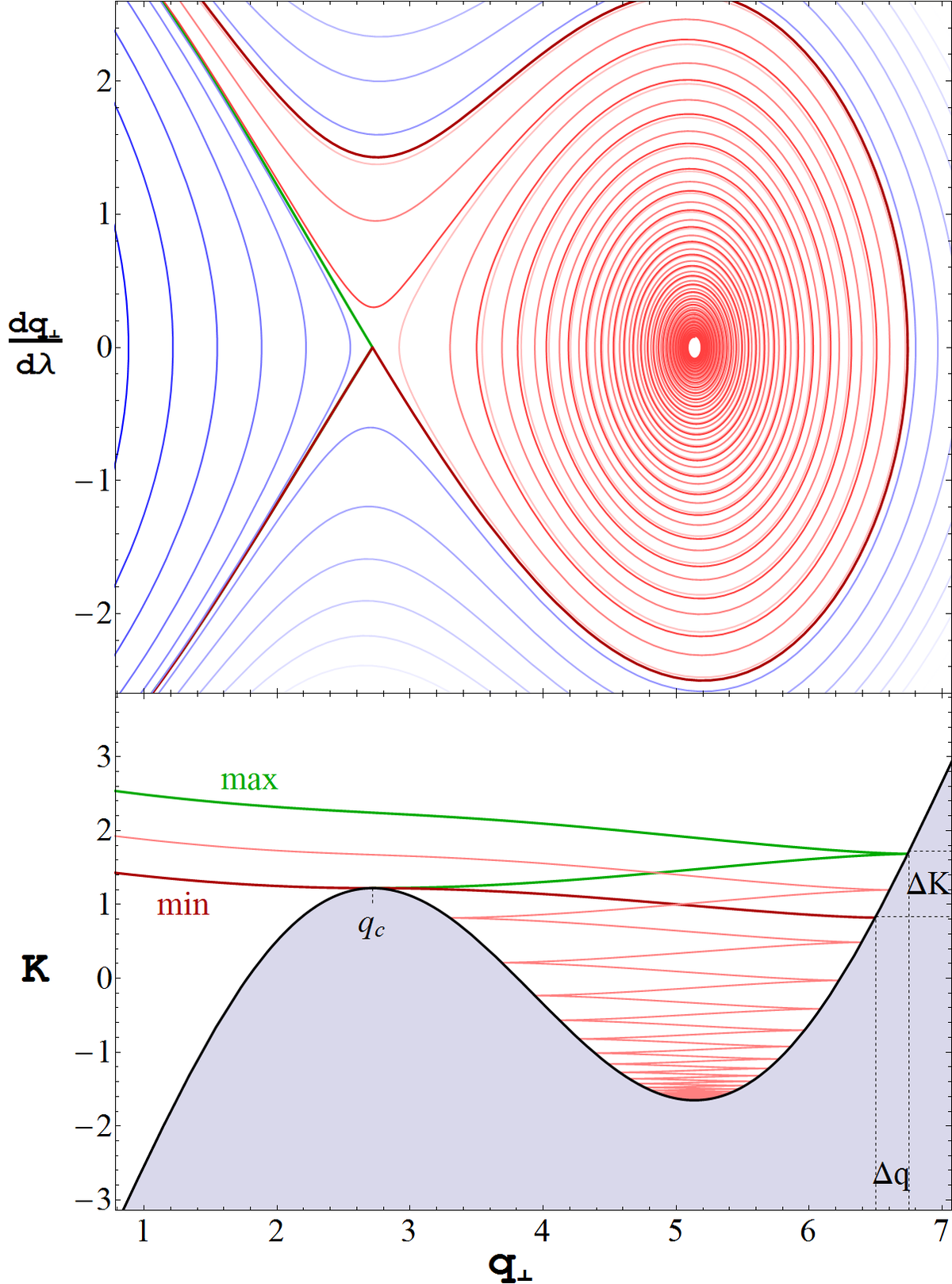}
\caption{Plotted are solutions of the same system as figure~\ref{fig:sust}, but with a constant dissipative term $\ei h^1_\perp =-1/15$ added. This allows solutions coming in from negative infinity to be captured in the local potential well.  The thick line labelled ``max''  is the solution with the highest effective energy  $\COM_{max}$ to be captured, while the thick  line labelled ``min'' corresponds to the solution with the minimal effective energy  $\COM_{min}$ that is still captured in the potential.}\label{fig:capture}
\end{figure}

From the expansion \eqref{eq:transexp}, we can find the next order contribution to the second order equation of motion for $\wip$,
\begin{equation}\label{eq:eom2}
\dd{\wip}{\mti}= -\d{(V)}{\wip}(\wip) + \ei \Ui_j^1 h_j^1(\wip)+\bigO(\ei^2),
\end{equation} 
with
\begin{align}
h^1_j(\wip)&=\pd{G_i}{\U_j}(\vec\Ua^0,\wip^0) + \pd{g_\perp}{\wip}(\vec\Ua^0,\wip)\delta_{\perp j} \\
&= \d{\Gca{\perp}}{\U_j}(\vec\Ua^0)+\\
&\quad\sum_{N\neq 0} \Bh{  \d{\Gc{\perp}{N}}{\U_j}(\vec\Ua^0)+\ii N \delta_{\perp j}\gc{\perp}{N}(\vec\Ua^0)}\ee^{\ii N \wip}.\notag
\end{align} 
The first-order $\ei$ correction couples the evolution of $\wip$ to that of the slowly evolving frequencies $\vec\Ui^1$ defined in \eqref{eq:transexp}. The $\vec\Ui^1$ correction acts as a friction term in the effective equations of motion for $\wip$, allowing solutions coming in from $\wip=-\infty$ to sink into a local minimum of the potential (see figure \ref{fig:capture}).

To make this notion precise, recall that the leading-order system had a constant of motion,
\begin{equation}
\COM  = \frac{1}{2}\Bh{\d{\wip}{\mti}}^2 + V(\wip).
\end{equation} 
When we include the higher-order corrections, this quantity is no longer constant. Taking a derivative and plugging in \eqref{eq:eom2} gives,
\begin{align}
\d{\COM}{\mti} &=\d{\wip}{\mti} \Bh{\dd{\wip}{\mti}+\d{V}{\wip}},\\
&= \ei\d{\wip}{\mti}  \Ui_j^1 h_j^1(\wip),\label{eq:Kode}
\end{align} 
which can be integrated to obtain $\COM$ as a function of $\mti$,
\begin{equation}\label{eq:Ksol}
\COM(\mti) = \COM_0 +\ei \int_{0}^{\mti} \d{\wip}{\mti} \Ui_j^1 h_j^1(\wip) \id{\mt}, 
\end{equation} 
where $\COM_0$ is the value of $\COM$ at $\mti=0$. (Recall that  $\mti=0$  is the time that the system (first) passes through resonance, i.e. $\Ui_\perp=0$ at $\mti=0$.)

In order for the system to be captured in sustained resonance, $\COM$ must decrease sufficiently as $\wip$ ``passes over'' the local minimum of the potential. In general, there will be a window of values $\COM_0\in[K_{min},K_{max}]$ for which the solution is captured. 

The lowest extreme, $\COM_{min}$ corresponds to the solution that comes in from negative infinity, barely scrapes the top of the local maximum of the potential, and then reflects of the potential at $V=\COM_{min}$. The solution corresponding to the highest extreme $\COM_{max}$, first reflects of the potential at $V=\COM_{max}$ and then asymptotically approaches the local maximum of the potential from the right.

To calculate the values of  $\COM_{min}$ and  $\COM_{max}$, we use the fact that these extremal solutions are close to the critical solutions of the lowest order system, i.e. the solutions of \eqref{eq:eom1st} with $\COM$ equal to the local maximum of the potential $\COM_c$. We can use this to find an approximation of $\d{\wip}{\mti}$ as a function of $\wip$,
\begin{equation}\label{eq:dwiapprox}
\d{\wip}{\mti} =\pm \sqrt{2(\COM_c-V(\wip))} +\bigO(\ei).
\end{equation} 
We can then use  \eqref{eq:eomtransfreq} to obtain an approximation for $\vec\Ui^1$. Plugging \eqref{eq:dwiapprox} into  \eqref{eq:eomtransfreq} yields,
\begin{equation}
\d{\Ui^1_i}{\mti}\d{\mti}{\wip} 
= \frac{\Gca{i} + \sum\limits_{N\neq 0} \Gc{i}{N}\ee^{\ii N \wip}}{\pm \sqrt{2(\COM_c-V(\wip))}} +\bigO(\ei),
\end{equation}
where here and in the rest of the equations in this section the functions  $\Gca{i}$ and $ \Gc{i}{N}$ (and their derivatives) are understood to be evaluated at the initial $\vec\Ua^0$. This equation can be integrated to obtain,
\begin{equation}
\Ui^1_i(\wip) = \int_{\wi_{0}}^{\wi_{\perp}} \frac{\Gca{i} + \sum\limits_{N\neq 0} \Gc{i}{N}\ee^{\ii N s}}{ \sqrt{2(\COM_c-V(s))}}\id{s} +\bigO(\ei).
\end{equation} 
We can then use the approximations for  $\d{\wip}{\mti}$ and $\vec\Ui^1$ to obtain  leading-order approximations for $\COM_{min}$ and $\COM_{max}$ from \eqref{eq:Ksol}
\begin{align}
\COM_c &= \COM_{min} -\ei \int_{-\infty}^{0} \d{\wip}{\mti}  \Ui_j^1 h_j^1(\wip) \id{\mt}+\bigO(\ei^2)\\
&= \COM_{min} -\ei \int_{\wi_c}^{\wi_0} \Ui_j^1(\wi) h_j^1(\wi) \id{\wi}+\bigO(\ei^2),
\end{align} 
and
\begin{align}
\COM_c &= \COM_{max} +\ei \int_{0}^{\infty} \d{\wip}{\mti}  \Ui_j^1 h_j^1(\wip) \id{\mt}+\bigO(\ei^2)\\
&= \COM_{max} -\ei \int_{\wi_0}^{\wi_c}  \Ui_j^1(\wi) h_j^1(\wi) \id{\wi}+\bigO(\ei^2).
\end{align} 
Together these give the size of the window $\Delta\COM=\COM_{max}- \COM_{min}$,
\begin{align}
\Delta\COM &= 2\ei \int_{\wi_0}^{\wi_c}  \Ui_j^1(\wi) h_j^1(\wi) \id{\wi}+\bigO(\ei^2)\\
\begin{split}
&=2\ei \sum_j\int_{\wi_0}^{\wi_c} \id{\wi}   \int_{\wi_{0}}^{\wi}\id{s}  \frac{\Gca{j} + \sum\limits_{N\neq 0} \Gc{j}{N}\ee^{\ii N s}}{ \sqrt{2(\COM_c-V(s))}}\\
&\qquad\quad\times\Bh{ 
	\d{\Gca{\perp}}{\U_j}
	+
	\sum\limits_{N\neq0}  \bh{  \d{\Gc{\perp}{N}}{\U_j}+
		\\
&\qquad\qquad\quad+ \ii N \delta_{\perp j} \gc{\perp}{N}}\ee^{\ii N \wi}} +\bigO(\ei^2).
\end{split}
\end{align} 
The first thing we notice is that the size of the window in $\COM$ is very small (of order $\ei=\e^{1/2}$).  So, if we have an EMRI system with $\e = 10^{-6}$ that has a resonance strong enough to have sustained resonance solutions, only about 1 in 1000 inspirals will be captured. The rest will simply experience a transient resonance and obtain a kick to their constants of motion given by  \eqref{eq:mtjump} and \eqref{eq:Ujump}. 

Since the window $\Delta\COM$ is small we may obtain the window for the initial resonant phase $\Delta\wi$ as,
\begin{equation}
\Delta\wi = V'(\wi_0)\Delta\COM.
\end{equation} 
Since $V' = \bigO(1)$, this implies that the window  $\Delta\wi$ is also small.

Given the existence of sustained resonance solutions, a necessary and sufficient condition for the existence of captured inspiralling solutions is that $\COM_{max}\geq \COM_{min}$. The intuitive interpretation of this condition is that the friction term in \eqref{eq:eom2} must act to decrease $\COM$ most of the time, i.e. if we heuristically  think  of $\Ui^1_j$ as $\d{\wip}{\mti}$, then  $h^1_j$ must be mostly negative. Given that along an inspiral we heuristically expect the Mino frequencies to decrease\footnote{Recall that, because Mino time is rescaled by a factor $\Sigma= r^2 + a^2\cos^2\theta$ with respect to proper time, the frequencies with respect to Mino time increase with increasing radius, unlike frequencies with respect to proper time and Boyer-Lindquist time.} and the self-force to increase, we do indeed in general expect   $h^1_j$ to  be negative. This is by no means guaranteed, but we are given a general indication of the sign of $\Delta\COM$. However, since the existence of sustained resonance solutions may require non-generic conditions---if they exist in EMRIs at all (see section \ref{sec:test})---non-generic behaviour of $h^1_j$  certainly is not excluded.

\subsection{Escape from sustained resonance}\label{sec:escape}
Once an inspiral has been captured in resonance it will continue to evolve according to the equations of motion~\eqref{eq:eomsust}, until it manages to escape from the resonance. In general, there are two ways in which the system can escape from sustained resonance. The first is the reverse of the capture condition, i.e. the friction term in \eqref{eq:eom2} should have the ``wrong'' (i.e. negative) sign for an extended period of time. As we argued above, this is not what one heuristically expects to happen.

The other possibility is that the local minimum of the effective potential $V$ disappears as the system continues to evolve. As can be gathered from the geometry of the phase diagram in figure \ref{fig:phasediag}, evolution of the system along a resonant surface while still decreasing the energy implies that it becomes more circular. We know that for circular orbits the self-force can only depend on the $\w_\theta$ phase. Consequently the resonant oscillating terms of self-force must vanish as the orbit approaches circularity, and the local minimum of the effective potential $V$ must disappear before the orbit becomes fully circular.

However, without a more detailed knowledge of the functional form of the self-force this is all we can say about the evolution of an inspiral captured in sustained resonance.

\section{Evaluating the resonance conditions}\label{sec:test}

In the previous sections we have derived two necessary and together sufficient conditions for the existence of EMRI solutions that are captured in sustained resonance at some frequencies $\vec\U$,
\begin{subequations}\label{eq:conds}
\begin{align}
\Delta G_\perp(\vec\U) \geq 1,\label{eq:cond1}\\
\Delta\COM(\vec\U) \geq 0.\label{eq:cond2}
\end{align} 
\end{subequations} 
Note that since the $(n,k)$-modes of the self-force are expected to  become small for large values of $n$ and $k$, one expects $\Delta G_\perp$ to become small for resonances with large $n_r$ and $n_\theta$. Consequently, one can at best expect \eqref{eq:cond1} to be satisfied for low order resonances (i.e. $\abs{n_r/n_\theta} = 2/3, 2/4, 2/5,\dots$).

A useful fact for testing  \eqref{eq:cond1} is that, as a result of the resonance condition, $\w_\perp$ is constant on any resonant orbit, and thus forms an addition constant of motion. This allows us to study the long term average value of $G_\perp$---which is related to $\nEd$, $\nLd$, and $\nQd$ through equation \eqref{eq:ELQtoU}---on geodesic orbits with different $\w_\perp$ to find the dependence of  $G_\perp$ on $\w_\perp$. This can then be used to obtain $\Delta G_\perp$. It is worth emphasizing that this procedure will provide all Fourier modes of  $G_\perp$ that depend only on $\w_\perp$, including those that are odd under the transformation $\w_\perp\mapsto2\pi-\w_\perp$, despite the fact that for non-resonant orbits these later modes are associated with the conservative self-force.
   
In previous works, the additional constant of motion present for resonant orbits is defined in different ways. In \cite{Isoyama:2013yor}, they define $\Delta\mt$ as Mino time difference between the minima of the $r$ and $\theta$ oscillations of the orbits. If we define the generalized angles $q_r$ and $q_\theta$ to be zero at their minima, then this implies
\begin{subequations}
\begin{align}
\w_r(\mt) &= (\mt-\mt_0) \U_r,\\
\w_\theta(\mt) &= (\mt-\mt_0+\Delta\mt) \U_\theta,
\end{align} 
\end{subequations}
where $\mt_0$ is the time at which the $r$ oscillations reach their minimum. Plugging this in the definition of $\w_\perp$ gives
\begin{subequations}
\begin{align}
\w_\perp(\mt) &= n_r w_r(\mt)+n_\theta w_\theta(\mt) \\
&= n_r  (\mt-\mt_0) \U_r)+n_\theta  (\mt-\mt_0+\Delta\mt) \U_\theta\\
&= n_\theta\U_\theta\Delta\lambda,
\end{align} 

\end{subequations} 
where in the last line we used the resonance condition $n_r \U_r+n_\theta \U_\theta=0$. Hence $\Delta\lambda = \w_\perp/(n_\theta\U_\theta)$.

In \cite{Flanagan:2012kg}, $\chi_0$ is defined as,
\begin{equation}
\chi_0 = \arccos\Bh{\frac{z(\mt_0)}{z_{max}}},
\end{equation} 
where $z=\cos\theta$. Although rather complicated, the relation between $\chi_0$ and $\w_\perp$ can be obtained explicitly from the analytic solutions of the geodesic equations in Kerr found in \cite{Fujita:2009bp}.

Since $\Delta G_\perp$ only depends on the minimum and maximum of $G_\perp$, it does not matter whether we obtain  $G_\perp$ as a function of $\w_\perp$, $\Delta\lambda$, or $\chi_0$. The extrema are always the same.

In \cite{Flanagan:2012kg}, $\dot{\vec{P}} =(\nEd,\nLd,\nQd)$ was calculated by solving the Teukolsky equation for resonant orbits, and obtaining the ``fluxes'' at infinity and the horizon. They find that $\Delta\dot{P}_i$ (defined analogous to the definition in \eqref{eq:deltaG} of $\Delta G_\perp$) is at most of the order of a few percent. Since they only give $\Delta\dot{P}_i$, and not the values of $\dot{P}_i$, we cannot calculate $\Delta G_\perp$ directly from their results, but unless $\dot{P}_i$ is almost tangent to the resonant surface $\Delta G_\perp$  should be of similar order of magnitude as $\Delta\dot{P}_i$. This result is consistent with the conclusions of Flanagan and Hinderer \cite{Flanagan:2010cd} based on a 'post-Newtonian approximation of the self-force (even though the resonant orbits are in the strong field regime where the PN approximations loose their validity). Such small values of $\Delta G_\perp$  indicate that there are no sustained resonance solutions (let alone captured sustained resonances) for the resonant orbits probed in~\cite{Flanagan:2012kg}. 

However \cite{Flanagan:2012kg} probes only a few points in the parameter space.  Consequently, we cannot quite exclude the possibility of sustained resonances.

Testing the second condition for capture, \eqref{eq:cond2}, requires not only knowledge of $G_i$, but also of its derivatives with respect to $\U_i$. Consequently, we cannot rely on the shortcut provided by calculating the changes of the constants of motion on resonant geodesics. To test it, a complete survey of the self-force in a neighbourhood of the resonant surface would be needed. Such a survey, is currently beyond the state-of-the-art.

\section{Discussion and Conclusions}
In this paper we derived a set of necessary and sufficient conditions \eqref{eq:conds} that the self-force needs to satisfy for sustained resonances to occur in extreme mass ratio inspirals. Along the way we obtained an expression (equations  \eqref{eq:mtjump} and \eqref{eq:Ujump}) for the jump made by the constants of motion when an EMRI encounters a transient resonance. This expression---valid to lowest order in the mass ratio $\e$---applies to resonances of any strength, including transient crossings of resonances strong enough to allow the existence of sustained resonances.

Current numerical  evidence provides no indication that condition  \eqref{eq:cond1} is satisfied for any resonant orbits in Kerr spacetime. However, these results are  not sufficient to completely rule out that the condition may be satisfied for some particular resonant orbits. Efforts to provide a more comprehensive sweep of the parameter space of resonant surfaces are currently under way. Numerical testing of the second condition \eqref{eq:cond2} would require almost full knowledge of the self-force for generic orbits in Kerr spacetime, which at this point is not (yet) available.

The initial conditions needed to allow capture into sustained resonance however imply that even if the conditions \eqref{eq:conds} are satisfied for some orbits, the probability that an EMRI crossing that orbit is indeed captured is only of order $\sim \e^{1/2}$. This makes the chance of actually observing such an event in an astrophysical context fairly remote, unless very large numbers of EMRIs are observed. However, if an EMRI stuck in sustained resonance is ever observed, it would provide a unique glance into the resonant structure of the space of orbits in Kerr, which as pointed out in \cite{Brink:2013nna}, is sensitive to the geometry of the spacetime.
\begin{acknowledgments}
The author would like to thank Leor Barack for many helpful discussions. He also likes to thank Leor Barack and Priscilla Ca\~nizares for a careful reading of this paper. This work was supported by a NWO Rubicon grant.
\end{acknowledgments}

\appendix
\section{Expressions for the forcing terms}\label{app:SF}
In \cite{Hinderer:2008dm} Hinderer and Flanagan derive the expression for the forcing terms $\tilde{g}_i$ and $\tilde{G}_i$ in terms of the self-acceleration $a_\nu$,
\begin{subequations}
\begin{align}
\tilde{g}_i(\vec{P},\vec\w) &= \d{\pt}{\mt}\left(\pd{\w_i}{p_\nu}\right)_{x}a_\nu\\
 \tilde{G}_i(\vec{P},\vec\w) &= \d{\pt}{\mt}\left(\pd{P_i}{p_\nu}\right)_{x}a_\nu ,
\end{align} 
\end{subequations}
where $p_\nu$ is the four-momentum, and the parentheses with subscript $x$ mean that the partial derivatives are performed keeping the position $x^\nu$ fixed. They also show how to calculate the partial derivatives more explicitly. The result for $\tilde{G}$ is rather simple,
\begin{align}
 \vec{\tilde{G}}(\vec{P},\vec\w) &= \d{\pt}{\mt}(-a_t,a_\phi,2 Q^{\mu\nu}u_\mu a_\nu).
\end{align} 
The expression for $\tilde{g}$ is much more involved, and since $\tilde{g}$ is not needed to evaluate the capture conditions derived in this paper, we will not repeat it here.

The expressions for the forcing terms for the equation of motion in terms of the frequencies, $G$ and $g$, can be easily derived from $\tilde{g}$ and $\tilde{G}$,
\begin{subequations}
\begin{align}
 G_i(\vec\U,\vec\w) &= \pd{\U_i}{P_j}\tilde{G}_j\big(\vec\U(\vec{P}),\vec\w\big),\label{eq:ELQtoU}\\
 g_j(\vec\U,\vec\w) &=\tilde{g}_j\big(\vec\U(\vec{P}),\vec\w\big).
\end{align} 
\end{subequations} 
Evaluating this requires knowledge  of $\vec\U(\vec{P})$. Unfortunately, no analytic form of $\vec\U(\vec{P})$ is known. However, it  is possible to explicitly get $\vec\U$ and $\vec{P}$ as functions of $\vec{s}=(p,e,z_{max})$\footnote{Here, $p$ is the semi-latus rectum, $e$ the eccentricity, and $z_{max}$ is the maximal value of $z=\cos\theta$.} (see  \cite{Schmidt:2002qk,Fujita:2009bp} for an explicit formula). We can therefore obtain $\vec\U(\vec{P})$ numerically  by numerically inverting the relation $\vec{p}(\vec{s})$. Similarly we obtain $\pd{\U_i}{P_j}$ from $\pd{\U_i}{s_j}$ and $\pd{P_i}{s_j}$,
\begin{equation}
\pd{\U_i}{P_j} = \pd{\U_i}{s_k}\pd{s_k}{P_j}.
\end{equation} 

\section{Carter-Mino time frequencies}\label{app:Minofreq}
In this paper we identify (invariant tori of) bound orbits by their frequencies with respect to Mino time, $\vec\U = (\U_r,\U_\theta, \U_\phi)$, instead of more conventional sets of invariants such as $(\nE,\nL,\nQ)$ or $(p,e,z_{max})$. This choice is convenient because it avoids the appearance of the function $\vec\U(\nE,\nL,\nQ)$ and its derivatives in many of the equations. However, this choice is not crucial for any of the conclusions of the paper.

For this choice to be valid, we need the map $F\colon(p,e,z_{max})\mapsto(\U_r,\U_\theta, \U_\phi)$ to be invertible, or at least for it be invertible in a neighbourhood of the resonant surfaces. It is known that this is not the case for the related map  $(p,e,z_{max})\mapsto(\Omega_r,\Omega_\theta, \Omega_\phi)$ to the frequencies with respect to Boyer-Lindquist coordinate time \cite{Warburton:2013yj}. In that case, there exist distinct invariant tori of bound orbits with the same triple of coordinate frequencies $(\Omega_r,\Omega_\theta, \Omega_\phi)$, and the region of parameter space where these isofrequency pairs appear intersects at least some of the low integer resonant surfaces.  This might be cause to worry that the map to Mino time frequency triples is equally degenerate.

We know of no formal proof that the map  $F$ is invertible, but our investigations indicate that this seems to be the case. We first present a formal proof that the reduced map $(p,e)\mapsto(\U_r,\U_\theta=\U_\phi)$ for orbits in Schwarzschild spacetimes is invertible, unlike the analogue map for $\Omega$. We then present numerical plots of the parameter space  for bound orbits in Kerr spacetime presented in Mino frequencies which show the map to be regular in the plotted region, which includes the low integer resonant surfaces.

\subsection{Schwarzschild case}
Due to the spherical symmetry of Schwarzschild spacetime the orbital dynamics are independent of the inclination of the orbit. Consequently, there are only two relevant invariant parameters for each orbit. The same symmetry also implies that the oscillations about the equatorial plane must match the azimuthal period of the orbit, i.e. $\U_\theta=\U_\phi$. This leaves a reduced map $F\colon (p,e)\mapsto(\U_r,\U_\theta=\U_\phi)$ to consider. The expressions for $\U_r,\U_\phi$ in terms of $(p,e)$ are readily obtained from the general expressions in \cite{Schmidt:2002qk,Fujita:2009bp},
\begin{subequations}
\begin{align}
\U_r &= \frac{\pi\sqrt{\frac{p(p-6+2e)}{p-e^2-3}}}{2 K(\frac{4e}{p-6+2e})},\\
\U_\phi &= \frac{p}{\sqrt{p-e^2-3}},
\end{align} 
\end{subequations} 
where
\begin{equation}
K(x) = \int_0^{\pi/2}\id{\sigma} \frac{1}{\sqrt{1-x\sin^2\sigma}}
\end{equation} 
is the complete elliptic integral of the first kind.

For convenience, we shift the semilatus rectum $p$ by the $6+2e$, the value of $p$ at the separatrix, i.e. $\ps=\frac{p-6+2e}{4e}$. As a consequence the space of bound orbits in Schwarzschild spacetime is given by $\ps\geq 1$ and $0\leq e<1$. In these variables the frequencies are
\begin{subequations}
\begin{align}
\U_r &= \frac{\pi\sqrt{\frac{e\ps(4e\ps-2e+6)}{4e\ps +(1-e)(3+e)}}}{ K(\frac{1}{\ps})},\\
\U_\phi &= \frac{4e\ps-2e+6}{\sqrt{4e\ps +(1-e)(3+e)}}.
\end{align} 
\end{subequations} 
Note that the map $(p,e)\mapsto (\ps,e)$ is invertible. Hence $F$ is invertible if and only if $\tilde F\colon (\ps,e)\mapsto (\U_r,\U_\phi)$ is invertible. By the inverse function theorem it is enough to show the Jacobian of $\tilde F$ is non-zero everywhere.

The Jacobian $J$ of the map $\tilde F$ is given by
\begin{multline}
J = 2e\pi\sqrt{e\ps(4e\ps-2e+6)}\times\\
\frac{2(\ps^2-\ps+1)E(\frac{1}{\ps})-(2\ps^2-3\ps+1)K(\frac{1}{\ps})}{(\ps-1)(4e\ps +(1-e)(3+e))^2K(\frac{1}{\ps})^2},
\end{multline}
where
\begin{equation}
E(x) = \int_0^{\pi/2}\id{\sigma} \sqrt{1-x\sin^2\sigma}
\end{equation} 
is the complete elliptic integral of the second kind.

Since the square and square root factors are manifestly positive, $J>0$ is equivalent to
\begin{equation}
f(\ps)=2(\ps^2-\ps+1)E(\frac{1}{\ps})-(2\ps^2-3\ps+1)K(\frac{1}{\ps})>0.
\end{equation} 
Note that $f$ is only a function of $\ps$ and not of $e$. We will show that $f$ is a monotonically increasing function $\ps$. Since $f(1)=2$, this would prove that $f$ is strictly positive. We first calculate the third derivative of $f$,
\begin{equation}
f'''= \frac{(2-\ps)E(\frac{1}{\ps})+(\ps-1) K(\frac{1}{\ps})}{8/15 (\ps-1)\ps^3}.
\end{equation} 
If we expand the elliptic integral in the numerator we find,
\begin{equation}
\int_0^{\pi/2}\id{\theta} (2-\ps)(1-\frac{\sin^2\theta}{1+\ps})^{1/2}+\frac{\ps-1}{ (1-\frac{\sin^2\theta}{1+\ps})^{1/2}}.
\end{equation} 
The integrand can be rewritten to,
\begin{equation}
\frac{\cos^2\theta +(1+\sin^2\theta)(\ps-1)}{\ps(1-\frac{\sin^2\theta}{\ps})^{1/2}}>0,
\end{equation} 
which is manifestly positive for all $\ps\geq 1$ and $0<\theta<\pi/2$. Consequently, $f'''>0$
for all $\ps\geq 0$.

The first and second derivative of $f$ are given by
\begin{align}
f' &= \frac{(8\ps^2-3\ps+2)E(\frac{1}{\ps})-(8\ps^2-7\ps-1) K(\frac{1}{\ps})}{2 \ps},\\
f'' &= \frac{(16\ps^2+4\ps+6)E(\frac{1}{\ps})-(16\ps^2-4\ps+3) K(\frac{1}{\ps})}{4 \ps^2}.
\end{align}
If we use that
\begin{align}
E(\frac{1}{\ps}) & = \frac{\pi}{2}(1 - \frac{1}{4\ps}-\frac{3}{64\ps^2}+ \bigO(\frac{1}{\ps^3}),\\
K(\frac{1}{\ps}) & = \frac{\pi}{2}(1 + \frac{1}{4\ps}+\frac{9}{64\ps^2}+ \bigO(\ps^3),
\end{align} 
we find both vanish at infinity,
\begin{align}
\lim_{\ps\to\infty} f' &= 0,\\
\lim_{\ps\to\infty} f'' &= 0.
\end{align} 
Consequently, since $f'''$ is positive and $\lim_{\ps\to\infty} f'' = 0$, $f''$ must be strictly negative. So, $f'$ is a monotonically decreasing function of $\ps$. Since this also vanishes at infinity, it must be strictly positive. Hence $f$ is a monotonically increasing function. As a consequence, $f$ and by extension $J$ are strictly positive, and the inverse function theorem therefore implies that $\tilde{F}$ (and thus $F$) is invertible.

\subsection{Plots for Kerr}
\begin{figure*}[tbp]
\begin{minipage}{.48\textwidth}\centering
 \includegraphics[width=\linewidth]{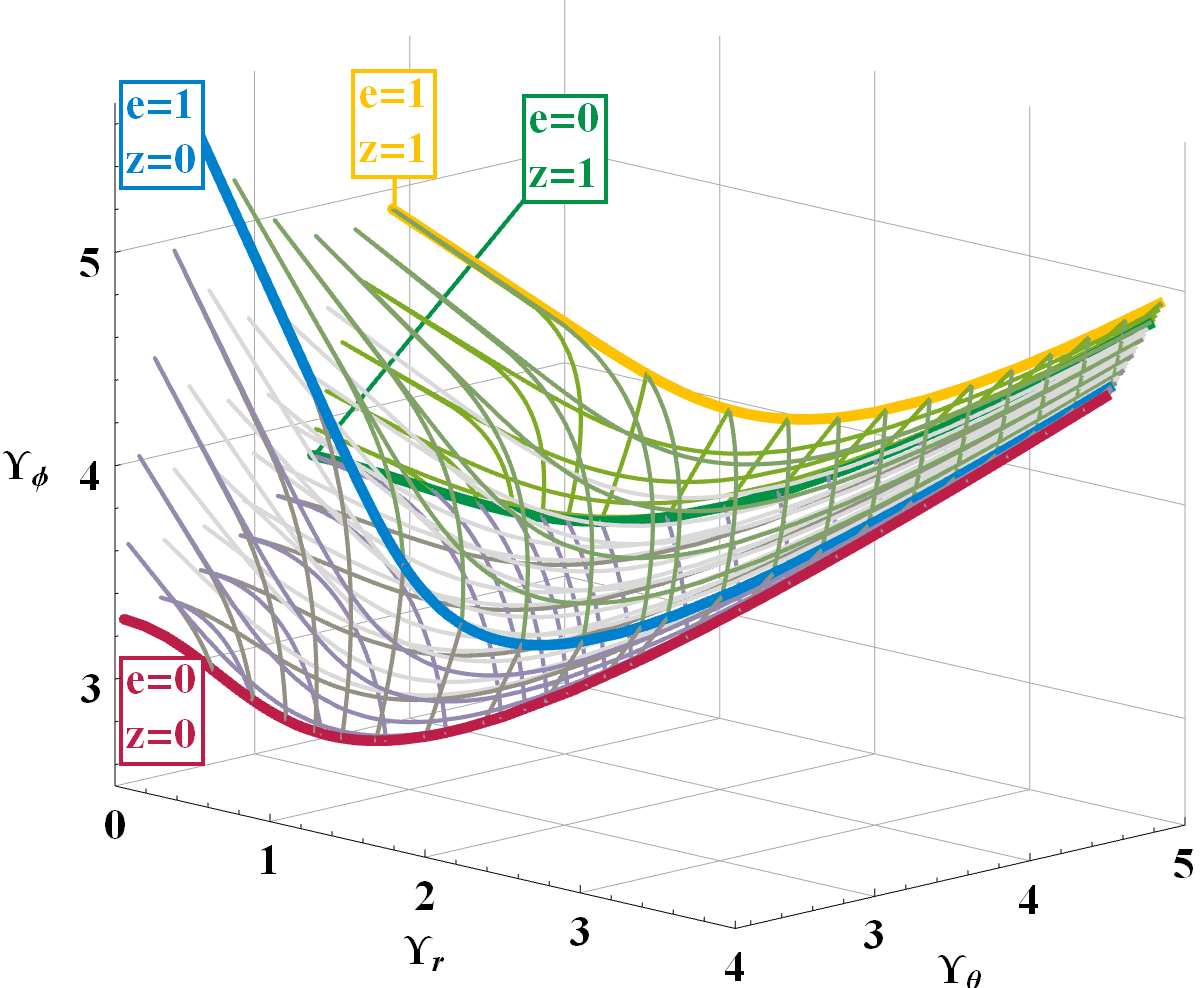}
\caption{Parameter space of prograde bound orbits in a Kerr space time with  $a=0.9$.}\label{fig:freqplot}
\end{minipage}\hspace{2em}%
\begin{minipage}{.48\textwidth}\centering
 \includegraphics[width=\linewidth]{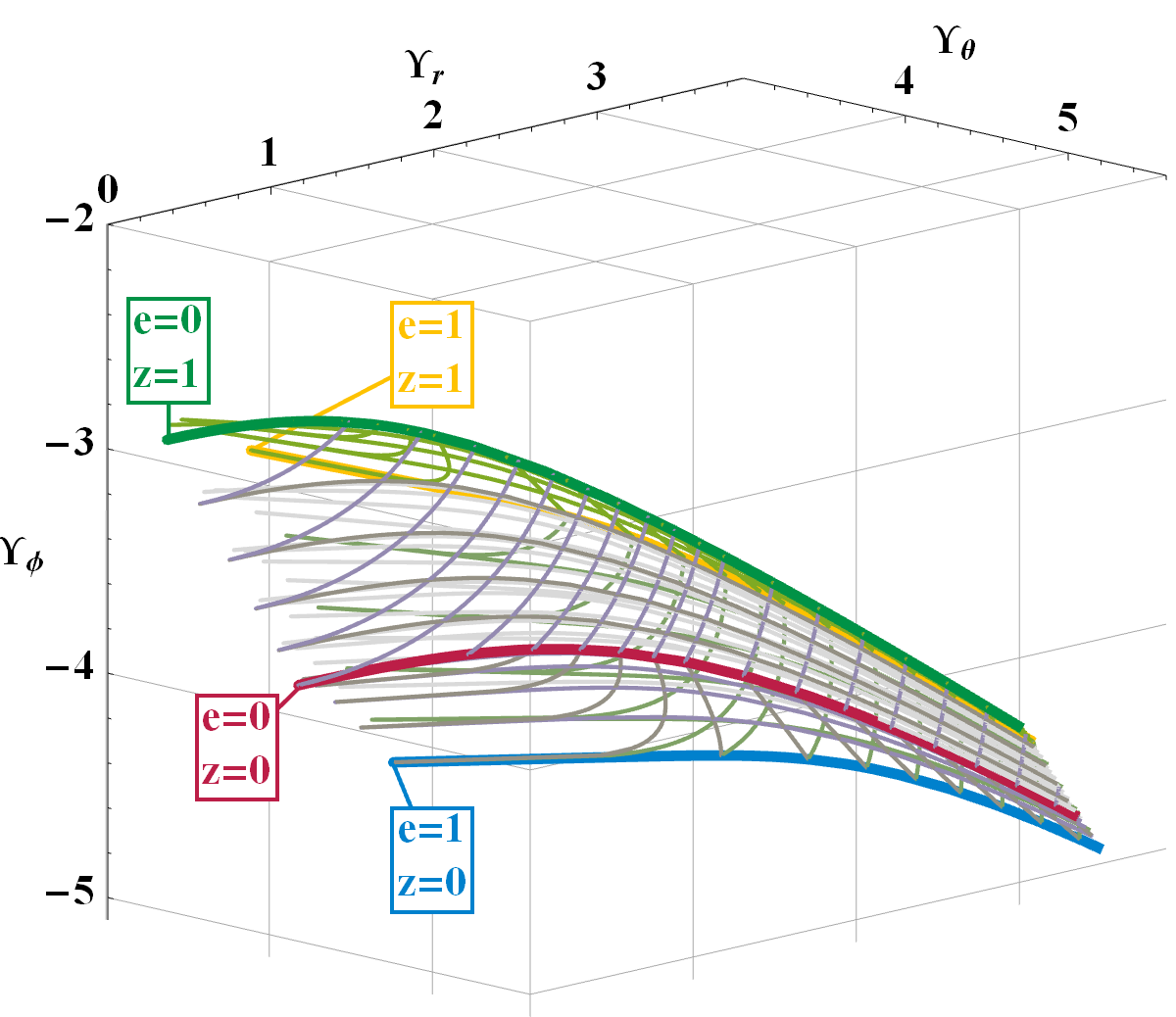}
\caption{The same parameter space as in figure \ref{fig:freqplot}, but for retrograde orbits.}\label{fig:retrofreqplot}
\end{minipage} 
\end{figure*} 
The general expressions for $(\U_r,\U_\theta,\U_\phi)$ in terms of $(p,e,z_{max})$ in a Kerr spacetime are known. \cite{Schmidt:2002qk,Fujita:2009bp} However, they are complicated enough that a direct computation of the Jacobian seems intractable, and we are not going to attempt it here. Instead we use the explicit expressions to numerically plot parameter space of bound orbits in Kerr spacetime as parametrized by the Mino frequencies.

Figures \ref{fig:freqplot} and \ref{fig:retrofreqplot} show this parameter space for respectively prograde and retrograde orbits in a Kerr spacetime with spin $a=\tfrac{9}{10}M$. The separatrix dividing bound and plunge orbits is given by the surface $\U_r=0$. The plotted lines in the plot keep two of the original parameters $(p,e,z_{max})$ constant.  A failure of invertibility of $F$, would manifest itself as this grid degenerating. A similar plot for the frequencies with respect to Boyer-Lindquist coordinate time, easily reveals the isofrequency region where the grid folds back on itself.

A particularly strong clue that helps us think that there may be no Mino time isofrequency pairs, is that parabolic ($e=1$) orbits have a finite radial frequency. Hence, the separatrix and the plane of parabolic orbits cannot get (partly) mapped into each other as happens for coordinate and proper time frequencies. This more or less restricts to possibility of Mino time isofrequency pairings to the interior of the plots in figures \ref{fig:freqplot} and \ref{fig:retrofreqplot} (where it also does not appear to occur).

The absence of any visible degeneration in these plots, is a strong indication that the Mino time frequencies are a good set of parameters for bound orbits in Kerr. At least, in the plotted region which contains the low integer ratios of $\U_r$ and $\U_\theta$. This, by no means, is a proof that there are no Mino isofrequency pairs of bound orbits in Kerr spacetimes, but it  is good enough for us to assume this as a conjecture.

\section{Near identity averaging transformation}\label{app:niat}
In this appendix we describe the details of the near-identity averaging transformation needed to remove the non-resonant oscillatory terms from the equations of motion \eqref{eq:eom}. It closely follows the procedure described in section 5.1 of \cite{KC:1996}. Recall the equations of motion  \eqref{eq:eom} for an EMRI system,

\begin{subequations}\label{eq:Aeom}
\begin{align}
\d{\U_i}{\mt} &= \e G_i(\vec{\U},\vec{\w}) + \bigO(\e^2),\\
\d{\w_j}{\mt} &= \U_j +  \e g_j(\vec{\U},\vec{\w}) + \bigO(\e^2),
\end{align}
\end{subequations} 
with
\begin{align}
\begin{split}
G_i(\vec\U,\vec{\w}) &= \Gca{i}(\vec\U) + \sum_{N\neq 0} \Gc{i}{N}(\vec\U)\ee^{\ii N \w_\perp}\\ 
&\quad\qquad+\sum_{(n,k)\in R} \Gc{i}{nk}(\vec\U) \ee^{\ii n \w_r+\ii k \w_\theta},
\end{split} \\
\begin{split}
g_j(\vec\U,\vec{\w}) &= \gca{j}(\vec\U) + \sum_{N\neq 0} \gc{j}{N}(\vec\U)\ee^{\ii N \w_\perp}\\ 
&\quad\qquad+\sum_{(n,k)\in R} \gc{j}{nk}(\vec\U) \ee^{\ii n \w_r+\ii k \w_\theta}.
\end{split} 
\end{align} 

We introduce a change of variables that is an identity transformation at leading-order (hence the term ``near-indentity''),
\begin{subequations}\label{eq:niat}
\begin{align}
\Ua_i(\vec\U,\vec{\w})  &= \U_i +\e T_i(\vec\U,\vec{\w}) +\bigO(\e^2),\\
\wa_j(\vec\U,\vec{\w})  &= \w_j +\e L_j(\vec\U,\vec{\w}) +\bigO(\e^2).
\end{align} 
\end{subequations} 
The inverse transformation is given by
\begin{subequations}\label{eq:iniat}
\begin{align}
\U_i(\vec\Ua,\vec{\wa}) &= \Ua_i -\e T_i(\vec\Ua,\vec{\wa}) +\bigO(\e^2),\\
\w_j(\vec\Ua,\vec{\wa}) &= \wa_j -\e L_j(\vec\Ua,\vec{\wa}) +\bigO(\e^2).
\end{align} 
\end{subequations} 

To obtain the equations of motion for the new variables $\vec\Ua$ and $\vec\wa$ we first differentiate \eqref{eq:niat} with respect to $\mt$,
\begin{subequations}
\begin{align}
\d{\Ua_i}{\mt} &= \d{\U_i}{\mt} +\e\cbB{ \d{T_i}{\U_k}\d{\U_k}{\mt}+ \d{T_i}{\w_k}\d{\w_k}{\mt}}+\bigO(\e^2),
 \\
\d{\wa_j}{\mt} &= \d{\w_j}{\mt} +\e\cbB{\d{L_j}{\U_k}\d{\U_k}{\mt}+\d{L_j}{\w_k}\d{\w_k}{\mt}}+\bigO(\e^2).
\end{align} 
\end{subequations} 
If we then substitute the equations of motion \eqref{eq:Aeom}, and use the inverse transformation \eqref{eq:iniat} to eliminate the dependence on $\vec\U$ and $\vec\w$, we obtain,
\begin{subequations}
\begin{align}
\d{\Ua_i}{\mt} &=\e\cbb{G_i(\vec{\U},\vec{\w})+\d{T_i}{\w_k}\U_k} +\bigO(\e^2),\\
\d{\wa_j}{\mt} &= \Ua_j  +\e\cbb{ g_j(\vec{\U},\vec{\w})+ \d{L_j}{\w_k}\U_k-T_j} +\bigO(\e^2).
\end{align} 
\end{subequations}
The idea is to use the freedom in the functions $T$ and $L$ to eliminate the lowest order non-resonant oscillatory terms from the equations of motion. That is we want to set,
\begin{subequations}
\begin{align}
 \d{T_i}{\w_k}\U_k &=-\sum_{(n,k)\in R} \Gc{i}{nk}(\vec\U) \ee^{\ii n \w_r+\ii k \w_\theta},\\
 \d{L_j}{\w_k}\U_k &=\tilde{T}_j-\sum_{(n,k)\in R} \gc{j}{nk}(\vec\U)\ee^{\ii n \w_r+\ii k \w_\theta},
\end{align} 
\end{subequations}
where $\tilde T_i$ denotes the terms of $T_i$ that depend on $\vec\w$. This set of first-order ODEs can easily be solved by direct integration, yielding
\begin{subequations}\label{eq:niatsol}
\begin{align}
T_i(\vec\U,\vec\w) &= \bar{T}_i(\vec\U)+\ii\sum_{\substack{(n,k)\\ \in R}}\frac{\Gc{i}{nk}(\vec\U)}{ n \U_r+ k \U_\theta} \ee^{\ii n \w_r+\ii k \w_\theta},\\
\begin{split}
L_j(\vec\U,\vec\w) &= \bar{L}_j(\vec\U)+\ii\sum_{\substack{(n,k)\\ \in R}}\frac{\gc{j}{nk}(\vec\U)}{ n \U_r+ k \U_\theta}\ee^{\ii n \w_r+\ii k \w_\theta}\\
&\qquad\quad+\sum_{\substack{(n,k)\\ \in R}}\frac{\Gc{j}{nk}(\vec\U)}{( n \U_r+ k \U_\theta)^2} \ee^{\ii n \w_r+\ii k \w_\theta}
\end{split}
\end{align}
and consequently
\begin{equation}
\tilde T_i(\vec\U,\vec\w) = \ii\sum_{(n,k)\in R}\frac{\Gc{i}{nk}(\vec\U)}{ n \U_r+ k \U_\theta} \ee^{\ii n \w_r+\ii k \w_\theta}.
\end{equation} 
\end{subequations}
The appearance of the combination $n \U_r+ k \U_\theta$ in the denominator of the solution, immediately tells why this procedure cannot be used to remove the resonant oscillatory terms, since these would become singular at resonance.

The solution\eqref{eq:niatsol} contains the arbitrary functions $\bar{T}_i$ and $\bar{L}_j$ of $\vec\U$. In principle, we could set them to zero, since our objective of removing the non-resonant oscillatory terms has been achieved. However, as explained in section 5.1 of \cite{KC:1996}, this freedom can be used to make further simplifications to the equations of motion. In \cite{KC:1996} this freedom is used to, eliminate the $\bigO(\e^2)$ averaged terms. We make a slightly different choice, we set $\bar{T}_j(\vec\U)=\gca{j}(\vec\U)$,which eliminates the $\bigO(\e)$ averaged term from the $\wa$ equations of motion, while we use  $\bar{L}_j(\vec\U)$ as in \cite{KC:1996} to eliminate the  $\bigO(\e^2)$ averaged terms from those same equations.

The equations of motion to  $\bigO(\e^2)$ resulting from these manipulations are.
\begin{subequations}
\begin{align}
\d{\Ua_i}{\mt} &= \e \Gca{i}(\vec\Ua) +\e \sum_{N\neq 0} \Gc{i}{N}(\vec\Ua)\ee^{\ii N \wa_\perp}+ \bigO(\e^2);\\
\d{\wa_j}{\mt} &= \Ua_j +\e \sum_{N\neq 0} \gc{j}{N}(\vec\Ua)\ee^{\ii N \wa_\perp} + \bigO(\e^2).
\end{align} 
\end{subequations} 

\bibliography{journalshortnames,susres}

\begin{thebibliography}{24}%
\makeatletter
\providecommand \@ifxundefined [1]{%
 \@ifx{#1\undefined}
}%
\providecommand \@ifnum [1]{%
 \ifnum #1\expandafter \@firstoftwo
 \else \expandafter \@secondoftwo
 \fi
}%
\providecommand \@ifx [1]{%
 \ifx #1\expandafter \@firstoftwo
 \else \expandafter \@secondoftwo
 \fi
}%
\providecommand \natexlab [1]{#1}%
\providecommand \enquote  [1]{``#1''}%
\providecommand \bibnamefont  [1]{#1}%
\providecommand \bibfnamefont [1]{#1}%
\providecommand \citenamefont [1]{#1}%
\providecommand \href@noop [0]{\@secondoftwo}%
\providecommand \href [0]{\begingroup \@sanitize@url \@href}%
\providecommand \@href[1]{\@@startlink{#1}\@@href}%
\providecommand \@@href[1]{\endgroup#1\@@endlink}%
\providecommand \@sanitize@url [0]{\catcode `\\12\catcode `\$12\catcode
  `\&12\catcode `\#12\catcode `\^12\catcode `\_12\catcode `\%12\relax}%
\providecommand \@@startlink[1]{}%
\providecommand \@@endlink[0]{}%
\providecommand \url  [0]{\begingroup\@sanitize@url \@url }%
\providecommand \@url [1]{\endgroup\@href {#1}{\urlprefix }}%
\providecommand \urlprefix  [0]{URL }%
\providecommand \Eprint [0]{\href }%
\providecommand \doibase [0]{http://dx.doi.org/}%
\providecommand \selectlanguage [0]{\@gobble}%
\providecommand \bibinfo  [0]{\@secondoftwo}%
\providecommand \bibfield  [0]{\@secondoftwo}%
\providecommand \translation [1]{[#1]}%
\providecommand \BibitemOpen [0]{}%
\providecommand \bibitemStop [0]{}%
\providecommand \bibitemNoStop [0]{.\EOS\space}%
\providecommand \EOS [0]{\spacefactor3000\relax}%
\providecommand \BibitemShut  [1]{\csname bibitem#1\endcsname}%
\let\auto@bib@innerbib\@empty
\bibitem [{\citenamefont {Gair}\ \emph {et~al.}(2004)\citenamefont {Gair},
  \citenamefont {Barack}, \citenamefont {Creighton}, \citenamefont {Cutler},
  \citenamefont {Larson} \emph {et~al.}}]{Gair:2004iv}%
  \BibitemOpen
  \bibfield  {author} {\bibinfo {author} {\bibfnamefont {J.~R.}\ \bibnamefont
  {Gair}}, \bibinfo {author} {\bibfnamefont {L.}~\bibnamefont {Barack}},
  \bibinfo {author} {\bibfnamefont {T.}~\bibnamefont {Creighton}}, \bibinfo
  {author} {\bibfnamefont {C.}~\bibnamefont {Cutler}}, \bibinfo {author}
  {\bibfnamefont {S.~L.}\ \bibnamefont {Larson}},  \emph {et~al.},\ }\href
  {\doibase 10.1088/0264-9381/21/20/003} {\bibfield  {journal} {\bibinfo
  {journal} {Classical Quant. Grav.}\ }\textbf {\bibinfo {volume} {21}},\
  \bibinfo {pages} {S1595} (\bibinfo {year} {2004})},\ \Eprint
  {http://arxiv.org/abs/gr-qc/0405137} {arXiv:gr-qc/0405137 [gr-qc]}
  \BibitemShut {NoStop}%
\bibitem [{\citenamefont {Gair}\ \emph {et~al.}(2013)\citenamefont {Gair},
  \citenamefont {Vallisneri}, \citenamefont {Larson},\ and\ \citenamefont
  {Baker}}]{Gair:2012nm}%
  \BibitemOpen
  \bibfield  {author} {\bibinfo {author} {\bibfnamefont {J.~R.}\ \bibnamefont
  {Gair}}, \bibinfo {author} {\bibfnamefont {M.}~\bibnamefont {Vallisneri}},
  \bibinfo {author} {\bibfnamefont {S.~L.}\ \bibnamefont {Larson}}, \ and\
  \bibinfo {author} {\bibfnamefont {J.~G.}\ \bibnamefont {Baker}},\ }\href@noop
  {} {\bibfield  {journal} {\bibinfo  {journal} {Living Rev. Rel.}\ }\textbf
  {\bibinfo {volume} {16}},\ \bibinfo {pages} {7} (\bibinfo {year} {2013})},\
  \Eprint {http://arxiv.org/abs/1212.5575} {arXiv:1212.5575 [gr-qc]}
  \BibitemShut {NoStop}%
\bibitem [{\citenamefont {Barack}\ and\ \citenamefont
  {Cutler}(2004)}]{Barack:2003fp}%
  \BibitemOpen
  \bibfield  {author} {\bibinfo {author} {\bibfnamefont {L.}~\bibnamefont
  {Barack}}\ and\ \bibinfo {author} {\bibfnamefont {C.}~\bibnamefont
  {Cutler}},\ }\href {\doibase 10.1103/PhysRevD.69.082005} {\bibfield
  {journal} {\bibinfo  {journal} {Phys. Rev. D}\ }\textbf {\bibinfo {volume}
  {69}},\ \bibinfo {pages} {082005} (\bibinfo {year} {2004})},\ \Eprint
  {http://arxiv.org/abs/gr-qc/0310125} {arXiv:gr-qc/0310125 [gr-qc]}
  \BibitemShut {NoStop}%
\bibitem [{\citenamefont {Barack}(2009)}]{Barack:2009ux}%
  \BibitemOpen
  \bibfield  {author} {\bibinfo {author} {\bibfnamefont {L.}~\bibnamefont
  {Barack}},\ }\href {\doibase 10.1088/0264-9381/26/21/213001} {\bibfield
  {journal} {\bibinfo  {journal} {Classical Quant. Grav.}\ }\textbf {\bibinfo
  {volume} {26}},\ \bibinfo {pages} {213001} (\bibinfo {year} {2009})},\
  \Eprint {http://arxiv.org/abs/0908.1664} {arXiv:0908.1664 [gr-qc]}
  \BibitemShut {NoStop}%
\bibitem [{\citenamefont {Poisson}\ \emph {et~al.}(2011)\citenamefont
  {Poisson}, \citenamefont {Pound},\ and\ \citenamefont
  {Vega}}]{Poisson:2011nh}%
  \BibitemOpen
  \bibfield  {author} {\bibinfo {author} {\bibfnamefont {E.}~\bibnamefont
  {Poisson}}, \bibinfo {author} {\bibfnamefont {A.}~\bibnamefont {Pound}}, \
  and\ \bibinfo {author} {\bibfnamefont {I.}~\bibnamefont {Vega}},\ }\href@noop
  {} {\bibfield  {journal} {\bibinfo  {journal} {Living Rev. Rel.}\ }\textbf
  {\bibinfo {volume} {14}},\ \bibinfo {pages} {7} (\bibinfo {year} {2011})},\
  \Eprint {http://arxiv.org/abs/1102.0529} {arXiv:1102.0529 [gr-qc]}
  \BibitemShut {NoStop}%
\bibitem [{\citenamefont {Flanagan}\ and\ \citenamefont
  {Hinderer}(2012)}]{Flanagan:2010cd}%
  \BibitemOpen
  \bibfield  {author} {\bibinfo {author} {\bibfnamefont {E.~E.}\ \bibnamefont
  {Flanagan}}\ and\ \bibinfo {author} {\bibfnamefont {T.}~\bibnamefont
  {Hinderer}},\ }\href {\doibase 10.1103/PhysRevLett.109.071102} {\bibfield
  {journal} {\bibinfo  {journal} {Phys. Rev. Lett.}\ }\textbf {\bibinfo
  {volume} {109}},\ \bibinfo {pages} {071102} (\bibinfo {year} {2012})},\
  \Eprint {http://arxiv.org/abs/1009.4923} {arXiv:1009.4923 [gr-qc]}
  \BibitemShut {NoStop}%
\bibitem [{\citenamefont {Mino}(2005)}]{Mino:2005an}%
  \BibitemOpen
  \bibfield  {author} {\bibinfo {author} {\bibfnamefont {Y.}~\bibnamefont
  {Mino}},\ }\href {\doibase 10.1143/PTP.113.733} {\bibfield  {journal}
  {\bibinfo  {journal} {Prog. Theor. Phys.}\ }\textbf {\bibinfo {volume}
  {113}},\ \bibinfo {pages} {733} (\bibinfo {year} {2005})},\ \Eprint
  {http://arxiv.org/abs/gr-qc/0506003} {arXiv:gr-qc/0506003 [gr-qc]}
  \BibitemShut {NoStop}%
\bibitem [{\citenamefont {Tanaka}(2006)}]{Tanaka:2005ue}%
  \BibitemOpen
  \bibfield  {author} {\bibinfo {author} {\bibfnamefont {T.}~\bibnamefont
  {Tanaka}},\ }\href {\doibase 10.1143/PTPS.163.120} {\bibfield  {journal}
  {\bibinfo  {journal} {Prog. Theor. Phys. Suppl.}\ }\textbf {\bibinfo {volume}
  {163}},\ \bibinfo {pages} {120} (\bibinfo {year} {2006})},\ \Eprint
  {http://arxiv.org/abs/gr-qc/0508114} {arXiv:gr-qc/0508114 [gr-qc]}
  \BibitemShut {NoStop}%
\bibitem [{\citenamefont {Grossman}\ \emph {et~al.}(2012)\citenamefont
  {Grossman}, \citenamefont {Levin},\ and\ \citenamefont
  {Perez-Giz}}]{Grossman:2011ps}%
  \BibitemOpen
  \bibfield  {author} {\bibinfo {author} {\bibfnamefont {R.}~\bibnamefont
  {Grossman}}, \bibinfo {author} {\bibfnamefont {J.}~\bibnamefont {Levin}}, \
  and\ \bibinfo {author} {\bibfnamefont {G.}~\bibnamefont {Perez-Giz}},\ }\href
  {\doibase 10.1103/PhysRevD.85.023012} {\bibfield  {journal} {\bibinfo
  {journal} {Phys. Rev. D}\ }\textbf {\bibinfo {volume} {85}},\ \bibinfo
  {pages} {023012} (\bibinfo {year} {2012})},\ \Eprint
  {http://arxiv.org/abs/1105.5811} {arXiv:1105.5811 [gr-qc]} \BibitemShut
  {NoStop}%
\bibitem [{\citenamefont {Gair}\ \emph {et~al.}(2012)\citenamefont {Gair},
  \citenamefont {Yunes},\ and\ \citenamefont {Bender}}]{Gair:2011mr}%
  \BibitemOpen
  \bibfield  {author} {\bibinfo {author} {\bibfnamefont {J.}~\bibnamefont
  {Gair}}, \bibinfo {author} {\bibfnamefont {N.}~\bibnamefont {Yunes}}, \ and\
  \bibinfo {author} {\bibfnamefont {C.~M.}\ \bibnamefont {Bender}},\ }\href
  {\doibase 10.1063/1.3691226} {\bibfield  {journal} {\bibinfo  {journal} {J.
  Math. Phys.}\ }\textbf {\bibinfo {volume} {53}},\ \bibinfo {pages} {032503}
  (\bibinfo {year} {2012})},\ \Eprint {http://arxiv.org/abs/1111.3605}
  {arXiv:1111.3605 [gr-qc]} \BibitemShut {NoStop}%
\bibitem [{\citenamefont {Flanagan}\ \emph {et~al.}(2014)\citenamefont
  {Flanagan}, \citenamefont {Hughes},\ and\ \citenamefont
  {Ruangsri}}]{Flanagan:2012kg}%
  \BibitemOpen
  \bibfield  {author} {\bibinfo {author} {\bibfnamefont {E.~E.}\ \bibnamefont
  {Flanagan}}, \bibinfo {author} {\bibfnamefont {S.~A.}\ \bibnamefont
  {Hughes}}, \ and\ \bibinfo {author} {\bibfnamefont {U.}~\bibnamefont
  {Ruangsri}},\ }\href {\doibase 10.1103/PhysRevD.89.084028} {\bibfield
  {journal} {\bibinfo  {journal} {Phys. Rev. D}\ }\textbf {\bibinfo {volume}
  {89}},\ \bibinfo {pages} {084028} (\bibinfo {year} {2014})},\ \Eprint
  {http://arxiv.org/abs/1208.3906} {arXiv:1208.3906 [gr-qc]} \BibitemShut
  {NoStop}%
\bibitem [{\citenamefont {Ruangsri}\ and\ \citenamefont
  {Hughes}(2014)}]{Ruangsri:2013hra}%
  \BibitemOpen
  \bibfield  {author} {\bibinfo {author} {\bibfnamefont {U.}~\bibnamefont
  {Ruangsri}}\ and\ \bibinfo {author} {\bibfnamefont {S.~A.}\ \bibnamefont
  {Hughes}},\ }\href {\doibase 10.1103/PhysRevD.89.084036} {\bibfield
  {journal} {\bibinfo  {journal} {Phys. Rev. D}\ }\textbf {\bibinfo {volume}
  {89}},\ \bibinfo {pages} {084036} (\bibinfo {year} {2014})},\ \Eprint
  {http://arxiv.org/abs/1307.6483} {arXiv:1307.6483 [gr-qc]} \BibitemShut
  {NoStop}%
\bibitem [{\citenamefont {Isoyama}\ \emph {et~al.}(2013)\citenamefont
  {Isoyama}, \citenamefont {Fujita}, \citenamefont {Nakano}, \citenamefont
  {Sago},\ and\ \citenamefont {Tanaka}}]{Isoyama:2013yor}%
  \BibitemOpen
  \bibfield  {author} {\bibinfo {author} {\bibfnamefont {S.}~\bibnamefont
  {Isoyama}}, \bibinfo {author} {\bibfnamefont {R.}~\bibnamefont {Fujita}},
  \bibinfo {author} {\bibfnamefont {H.}~\bibnamefont {Nakano}}, \bibinfo
  {author} {\bibfnamefont {N.}~\bibnamefont {Sago}}, \ and\ \bibinfo {author}
  {\bibfnamefont {T.}~\bibnamefont {Tanaka}},\ }\href {\doibase
  10.1093/ptep/ptt034} {\bibfield  {journal} {\bibinfo  {journal} {Prog. Theor.
  Exp. Phys.}\ }\textbf {\bibinfo {volume} {2013}},\ \bibinfo {pages} {063E01}
  (\bibinfo {year} {2013})},\ \Eprint {http://arxiv.org/abs/1302.4035}
  {arXiv:1302.4035 [gr-qc]} \BibitemShut {NoStop}%
\bibitem [{\citenamefont {Brink}\ \emph {et~al.}(2013)\citenamefont {Brink},
  \citenamefont {Geyer},\ and\ \citenamefont {Hinderer}}]{Brink:2013nna}%
  \BibitemOpen
  \bibfield  {author} {\bibinfo {author} {\bibfnamefont {J.}~\bibnamefont
  {Brink}}, \bibinfo {author} {\bibfnamefont {M.}~\bibnamefont {Geyer}}, \ and\
  \bibinfo {author} {\bibfnamefont {T.}~\bibnamefont {Hinderer}},\ }\href@noop
  {} {\  (\bibinfo {year} {2013})},\ \Eprint {http://arxiv.org/abs/1304.0330}
  {arXiv:1304.0330 [gr-qc]} \BibitemShut {NoStop}%
\bibitem [{\citenamefont {Carter}(1968)}]{Carter:1968rr}%
  \BibitemOpen
  \bibfield  {author} {\bibinfo {author} {\bibfnamefont {B.}~\bibnamefont
  {Carter}},\ }\href {\doibase 10.1103/PhysRev.174.1559} {\bibfield  {journal}
  {\bibinfo  {journal} {Phys. Rev.}\ }\textbf {\bibinfo {volume} {174}},\
  \bibinfo {pages} {1559} (\bibinfo {year} {1968})}\BibitemShut {NoStop}%
\bibitem [{\citenamefont {Mino}(2003)}]{Mino:2003yg}%
  \BibitemOpen
  \bibfield  {author} {\bibinfo {author} {\bibfnamefont {Y.}~\bibnamefont
  {Mino}},\ }\href {\doibase 10.1103/PhysRevD.67.084027} {\bibfield  {journal}
  {\bibinfo  {journal} {Phys. Rev. D}\ }\textbf {\bibinfo {volume} {67}},\
  \bibinfo {pages} {084027} (\bibinfo {year} {2003})},\ \Eprint
  {http://arxiv.org/abs/gr-qc/0302075} {arXiv:gr-qc/0302075 [gr-qc]}
  \BibitemShut {NoStop}%
\bibitem [{\citenamefont {Hinderer}\ and\ \citenamefont
  {Flanagan}(2008)}]{Hinderer:2008dm}%
  \BibitemOpen
  \bibfield  {author} {\bibinfo {author} {\bibfnamefont {T.}~\bibnamefont
  {Hinderer}}\ and\ \bibinfo {author} {\bibfnamefont {E.~E.}\ \bibnamefont
  {Flanagan}},\ }\href {\doibase 10.1103/PhysRevD.78.064028} {\bibfield
  {journal} {\bibinfo  {journal} {Phys. Rev. D}\ }\textbf {\bibinfo {volume}
  {D78}},\ \bibinfo {pages} {064028} (\bibinfo {year} {2008})},\ \Eprint
  {http://arxiv.org/abs/0805.3337} {arXiv:0805.3337 [gr-qc]} \BibitemShut
  {NoStop}%
\bibitem [{\citenamefont {Schmidt}(2002)}]{Schmidt:2002qk}%
  \BibitemOpen
  \bibfield  {author} {\bibinfo {author} {\bibfnamefont {W.}~\bibnamefont
  {Schmidt}},\ }\href {\doibase 10.1088/0264-9381/19/10/314} {\bibfield
  {journal} {\bibinfo  {journal} {Classical Quant. Grav.}\ }\textbf {\bibinfo
  {volume} {19}},\ \bibinfo {pages} {2743} (\bibinfo {year} {2002})},\ \Eprint
  {http://arxiv.org/abs/gr-qc/0202090} {arXiv:gr-qc/0202090 [gr-qc]}
  \BibitemShut {NoStop}%
\bibitem [{\citenamefont {Fujita}\ and\ \citenamefont
  {Hikida}(2009)}]{Fujita:2009bp}%
  \BibitemOpen
  \bibfield  {author} {\bibinfo {author} {\bibfnamefont {R.}~\bibnamefont
  {Fujita}}\ and\ \bibinfo {author} {\bibfnamefont {W.}~\bibnamefont
  {Hikida}},\ }\href {\doibase 10.1088/0264-9381/26/13/135002} {\bibfield
  {journal} {\bibinfo  {journal} {Classical Quant. Grav.}\ }\textbf {\bibinfo
  {volume} {26}},\ \bibinfo {pages} {135002} (\bibinfo {year} {2009})},\
  \Eprint {http://arxiv.org/abs/0906.1420} {arXiv:0906.1420 [gr-qc]}
  \BibitemShut {NoStop}%
\bibitem [{\citenamefont {Warburton}\ \emph {et~al.}(2013)\citenamefont
  {Warburton}, \citenamefont {Barack},\ and\ \citenamefont
  {Sago}}]{Warburton:2013yj}%
  \BibitemOpen
  \bibfield  {author} {\bibinfo {author} {\bibfnamefont {N.}~\bibnamefont
  {Warburton}}, \bibinfo {author} {\bibfnamefont {L.}~\bibnamefont {Barack}}, \
  and\ \bibinfo {author} {\bibfnamefont {N.}~\bibnamefont {Sago}},\ }\href
  {\doibase 10.1103/PhysRevD.87.084012} {\bibfield  {journal} {\bibinfo
  {journal} {Phys. Rev. D}\ }\textbf {\bibinfo {volume} {87}},\ \bibinfo
  {pages} {084012} (\bibinfo {year} {2013})},\ \Eprint
  {http://arxiv.org/abs/1301.3918} {arXiv:1301.3918 [gr-qc]} \BibitemShut
  {NoStop}%
\bibitem [{\citenamefont {Kevorkian}\ and\ \citenamefont
  {Cole}(1996)}]{KC:1996}%
  \BibitemOpen
  \bibfield  {author} {\bibinfo {author} {\bibfnamefont {J.}~\bibnamefont
  {Kevorkian}}\ and\ \bibinfo {author} {\bibfnamefont {J.~D.}\ \bibnamefont
  {Cole}},\ }\href@noop {} {\emph {\bibinfo {title} {Multiple scale and
  singular perturbation methods}}},\ Vol.\ \bibinfo {volume} {114}\ (\bibinfo
  {publisher} {Springer New York},\ \bibinfo {year} {1996})\BibitemShut
  {NoStop}%
\bibitem [{\citenamefont {Hinderer}\ and\ \citenamefont {Flanagan}()}]{FH:??}%
  \BibitemOpen
  \bibfield  {author} {\bibinfo {author} {\bibfnamefont {T.}~\bibnamefont
  {Hinderer}}\ and\ \bibinfo {author} {\bibfnamefont {E.~E.}\ \bibnamefont
  {Flanagan}},\ }\href@noop {} {\enquote {\bibinfo {title} {{Two timescale
  analysis of extreme mass ratio inspirals in Kerr. II. Transient
  Resonances}},}\ }\bibinfo {note} {(unpublished)}\BibitemShut {NoStop}%
\bibitem [{\citenamefont {Bosley}\ and\ \citenamefont
  {Kevorkian}(1995)}]{Bosley:1995}%
  \BibitemOpen
  \bibfield  {author} {\bibinfo {author} {\bibfnamefont {D.~L.}\ \bibnamefont
  {Bosley}}\ and\ \bibinfo {author} {\bibfnamefont {J.}~\bibnamefont
  {Kevorkian}},\ }\href@noop {} {\bibfield  {journal} {\bibinfo  {journal}
  {Stud. Appl. Math.}\ }\textbf {\bibinfo {volume} {94}},\ \bibinfo {pages}
  {83} (\bibinfo {year} {1995})}\BibitemShut {NoStop}%
\bibitem [{\citenamefont {Haberman}(1983)}]{Haberman:1983}%
  \BibitemOpen
  \bibfield  {author} {\bibinfo {author} {\bibfnamefont {R.}~\bibnamefont
  {Haberman}},\ }\href {\doibase 10.1137/0143016} {\bibfield  {journal}
  {\bibinfo  {journal} {SIAM J. Appl. Math.}\ }\textbf {\bibinfo {volume}
  {43}},\ \bibinfo {pages} {244} (\bibinfo {year} {1983})}\BibitemShut
  {NoStop}%
\end{thebibliography}%

\end{document}